\documentclass[12pt,aps,prd,tightenlines,nofootinbib,floatfix]{revtex4-2}
\usepackage{latexsym,amsmath,amssymb,amsbsy,graphicx}
\usepackage{xcolor}

\begin{document}
	
\title{Exclusive semileptonic $B$ decays  to the ground and  excited states of  light mesons}

\author{V.\,O. \surname{Galkin} }
\email{vgalkin@frccsc.ru}

\affiliation{
	Federal Research Center ``Computer Science and Control'' Russian Academy of Sciences, Vavilov Street 40, Moscow, 119333 Russia}

\author{Xian-Wei \surname{Kang}}
\email{xwkang@bnu.edu.cn}
\affiliation{Key Laboratory of Beam Technology of the Ministry of Education,
School of Physics and Astronomy, Beijing Normal University, Beijing 100875, China}

	
\begin{abstract}
Exclusive semileptonic decays of $B$ mesons to the ground, radially and orbitally excited states of light mesons are investigated in the framework of the relativistic quark model based on the quasipotential approach. Such decays are very important for the determination of the Cabibbo-Kobayashi-Maskawa matrix element $|V_{ub}|$ and testing the nature of the excited light mesons. The mixing schemes for the isoscalar light mesons are considered. The possible assignment of the experimentally observed excited light mesons to the particular states is discussed. The form factors parameterizing weak decay matrix elements between the initial $B$ meson and final light mesons are calculated with the complete account of the relativistic effects, including contributions of the intermediate negative energy states and relativistic transformations of the meson wave functions from rest to the moving reference frame. As a result the transferred momentum squared, $q^2$, dependence of the form factors is explicitly determined in the whole accessible kinematical range. On the basis of the helicity formalism the decay branching fractions and asymmetry and polarization parameters are calculated. The obtained results for the semileptonic $B$ decays to the ground states of light mesons are confronted with available experimental data. On this basis the value of the  $|V_{ub}|$ matrix element is determined. It is found in good agreement with the value extracted from inclusive decays. Predictions for the branching fractions of the semileptonic $B$ decays to radially ($2S$ and $3S$) and orbitally ($1P$ and $2P$) excited light mesons are given. Comparison with the previous calculations is performed. It is found that several decays to the excited light mesons ($B\to\rho(1450)l\nu_l$, $B\to b_1(1235)l\nu_l$, $B\to a_1(1260)l\nu_l$, $B\to h_1(1170)l\nu_l$, $B\to a_2(1320)l\nu_l$, $B\to a_2(1700)l\nu_l$) have branching fractions of the order $10^{-4}$ and, thus, they can be measured at existing and future $B$ factories.

\end{abstract}

\maketitle	
	
\section{Introduction}
		
The semileptonic $B$ meson decays to light mesons are the important source of the determination of the Cabibbo-Kobayashi-Maskawa (CKM) \cite{Cab,KM} matrix element $|V_{ub}|$ which is one of the fundamental parameters of the Standard Model. Since the decay rate of the semileptonic decay is proportional to $|V_{ub}|^2$, the corresponding measurement will provide its value. There are two methods of the $|V_{ub}|$ extraction from such decays. The first one is the inclusive method ($B\to X_u l\nu_l$), in which no specific final light hadron is reconstructed and the measurement is performed for the sum of the final hadronic states. Such measurements  are very challenging because of the background from the CKM favored decays involving charmed hadrons that require  cuts to suppress the dominant background contributions. The theoretical description requires paramterizations of the shape functions, which take these cuts into account (for recent review see \cite{fpv} and references therein). The second one is the exclusive method, in which the specific final state is reconstructed. The theoretical description of the exclusive decays requires calculation of the hadronic matrix elements of the weak current, which are parameterized by the set of the invariant form factors. The experimental and theoretical uncertainties are comparable in these methods. However, the values of $|V_{ub}|$ extracted from inclusive and exclusive data currently differ by approximately 3 standard deviations \cite{hflav}. Another attempt to solve this discrepancy was done previously by us in Ref.~\cite{Kang2013}.

The theoretical investigation of the exclusive semileptonic $B$ decays to light mesons requires comprehensive understanding of the light meson spectroscopy. This is especially important for the excited light meson states. At present, a lot of experimental data is available on light mesons \cite{ParticleDataGroup:2024cfk}. The main problem consists in prescribing the observed meson to the corresponding quark model state. The situation is complicated by the presence of the possible tetraquarks, glueballs and hybrid mesons. Note that they can mix with mesons having the same quantum numbers. Such mixing is particularly important for the isosinglet mesons for which $q\bar q$ ($q=u,d$) and $s\bar s$ states mix \cite{ParticleDataGroup:2024cfk,jrp}. At present, the firm understanding of such mixings is missing. The predicted decay rates of the exclusive semileptonic $B$ decays to light mesons are sensitive to their quantum numbers and compositions. Therefore, the measurement of such decays can provide a valuable information about the nature of excited light mesons.

In this paper, we apply the relativistic quark model (RQM) based on the quasipotential approach \cite{Ebert2002} for study of the exclusive semileptonic $B$ mesons decays to light mesons. Decays to the ground state, radially and orbitally excited light mesons are investigated. The corresponding decays of $D$ mesons to the ground and excited states of light mesons were considered previously \cite{FGK,gs2025}. Note that some of the $B$ meson semileptonic decays were already investigated in RQM \cite{Ebert2006,Ebert2011,Faustov2022}. Here we significantly improve the previous results by more careful treatment of the meson mixings, considering significantly more excited states and presenting predictions for more observables. In the semileptonic decays the lepton part factorizes and it is easily calculated by standard methods. The hadronic part is represented by the matrix element of the flavor-changing charged current between meson states. This matrix element is parameterized by the set of the invariant form factors. We calculate these form factors in the framework of the quasipotential approach with the comprehensive treatment of the relativistic effects. They include relativistic contributions of the intermediate negative-energy states and relativistic transformations of meson wave functions from rest to the moving reference frame. All relativistic contributions are taken without application of any expansion. As a result, the form factors are expressed through the overlap integrals of the meson wave functions, and their momentum transfer squared, $q^2$, dependence is explicitly determined in the whole accessible kinematical range. This is very important for the considered decays since the $q^2$ range for the $B$ decays to light mesons is very broad. The meson wave functions were previously obtained in the $B$ meson \cite{Ebert2009} and light meson \cite{Ebert2009l} mass spectra calculations. On the basis of the calculated form factors, the total and differential decay rates are calculated. Different observable asymmetry and polarization parameters are calculated.

The paper is organized as follows. In Sec.~\ref{RQM} we briefly describe our relativistic quark model. Masses and wave functions of light mesons are discussed in Sec.~\ref{sec:mass}. Particular attention is paid for the consideration of meson mixing and assigning quark model states to the observed mesons. Calculation of the weak decay matrix elements is outlined in Sec.~\ref{sec:weak}.  The obtained values of the semileptonic decay form factors are given in Sec.~\ref{sec:ff}. The necessary formulas for calculation of the semileptonic total and differential decay rates as well as asymmetry and polarization parameters are presented in Sec.~\ref{sec:sem}. Section~\ref{sec:results} contains our results and their discussion: First, we extract the value of the CKM matrix element $|V_{ub}|$ from the available experimental data of semileptonic $B$ decays to the ground states of light mesons. Then we use this value to calculate the decay branching fractions. We compare our results with experimental data and previous calculations. Finally, Sec.~\ref{sec:concl} contains our conclusions.

\section{Relativistic quark model}
\label{RQM}
All calculations presented in this paper are performed in the relativistic quark model (RQM) based on the quasipotential approach \cite{Ebert2002}. Heavy and light mesons are described by the wave function $\Psi_M({\bf p})$ which satisfies the quasipotential equation \cite{Ebert2002}
\begin{equation}
\label{eq:quas}
{\left(\frac{b^2(M)}{2\mu_{R}}-\frac{{\bf
p}^2}{2\mu_R}\right)\Psi_{M}({\bf p})} =\int\frac{d^3 q}{(2\pi)^3}
 V({\bf p,q};M)\Psi_{M}({\bf q}),
\end{equation}
where ${\bf p}$ is the relative momentum of the quark and antiquark,
\[b^2(M)
=\frac{[M^2-(m_1+m_2)^2][M^2-(m_1-m_2)^2]}{4M^2},\]
is the relative momentum squared on the mass shell in the center of mass system, and
\[\mu_R=\frac{M^4-(m^2_1-m^2_2)^2}{4M^3}\]
is the relativistic reduced mass. Here $M$ is the meson mass and $m_{1,2}$ are quark masses. The left hand side of the equation (\ref{eq:quas}) contains the relativistic kinematics, which leads to the dependence of the $b^2(M)$ and $\mu_R$ on the meson mass $M$.

The relativistic dynamics is contained in the kernel of the equation $V({\bf p,q};M)$. This is the quasipotential of the quark-antiquark interaction in meson and it is constructed by the off-mass-shell scattering amplitude projected on the positive energy states. It includes the one-gluon exchange potential, which dominates at small distances, and the long-range  potential, which is the mixture of the Lorentz-vector and scalar linear confining terms and dominates at large distances. The resulting potential in the nonrelativistic limit reduces to the Cornell potential, and, thus, represents its relativistic generalization. It includes both spin-independent and spin-dependent terms, which explicitly emerge from the considered amplitude \cite{Ebert2009,Ebert2009l}. It is important to point out that we do not employ the expansion in the quark velocities and treat all relativistic contributions nonperturbatevely. This is especially important for the light quarks. As a result the complicated dependence on the meson mass in the quasipotential occurs. The expressions for the quasipotential are given in Ref.~\cite{Ebert2009l}.

All parameters of the model were fixed previously from the consideration of meson mass spectra and radiative decays \citep{Ebert2002}. { Although these parameters were fixed almost 25 years ago, there is no necessity to readjust them, since most results of the model agree well with the latest experimental data. It should be noted that many predictions have subsequently been confirmed experimentally, e.g., the masses of excited states of the $B_c$ meson, the masses of the doubly charmed baryons and excited states of heavy baryons, etc.    
}

\section{Masses and wave functions of the light  mesons} \label{sec:mass}

The mass spectra of heavy-light and light mesons were calculated in Refs. \cite{Ebert2009,Ebert2009l}. The meson masses and wave functions were obtained as the result of the numerical solution of the quasipotential equation (\ref{eq:quas}).  These masses are given in Table~\ref{Mass1}. Let us point out that RQM results \cite{Ebert2009} are in  agreement with experimental data \cite{ParticleDataGroup:2024cfk} and exhibit linear Regge trajectories. Even the mass of the $\pi$ has a reasonable value. This is the consequence of the completely relativistic treatment of the quark-antiquark kinematics and interaction. As it has been already pointed out, in our approach the nonlinear dependence on the meson mass occurs in the interaction quasipotentail. The term responsible for the spin-spin interaction contains the product of the on-mass-shell energies of the light quark and antiquark ($E_{1,2}=(M^2-m_{2,1}^2+m_{1,2}^2)/2M$)  in the denominator. For mesons containing a quark and antiquark  with the same constituent mass ($m_1=m_2$) these energies $E_{1,2}$ are equal to the half of the meson mass ($M/2$). As a result the spin-spin splitting is significantly increased for the lightest $\pi$ meson, correctly reproducing hyperfine splitting between $\pi$ and $\rho$. The masses of the isospin triplet ($I=1$) radially excited pseudoscalar and vector mesons as well as isospin singlet ($I=0$) vector mesons $\omega$ and $\phi$ are also well reproduced. We find that $\pi(1300)$, $\rho(1450)$, $\omega(1420)$ and $\phi(1680)$ can be considered as the first radial ($2S$) excitations, while $\pi(1800)$, $\rho(1960)$ and $\phi(2170)$ belong to the second radial ($3S$) excitation.

\begin{table}
	\caption{Calculated and experimental masses of light mesons (in MeV).}\label{Mass1}
	\begin{ruledtabular}
		\begin{tabular}{@{}ccc@{\!\!}c@{\!}c@{\!\!}c@{\!\!}c@{\!}c@{}c@{\!\!}c@{\!\!}c@{}}
		&&&\multicolumn{2}{c}{Mass}&&\multicolumn{2}{c}{Mass}&&\multicolumn{2}{c}{Mass}\\
		\cline{4-5} \cline{7-8} \cline{10-11}
		State&$J^{PC}$&$I=1$&RQM\cite{Ebert2009l}&Exp.\cite{ParticleDataGroup:2024cfk}&$I=0$&RQM\cite{Ebert2009l}&Exp.\cite{ParticleDataGroup:2024cfk}&$I=0$&RQM\cite{Ebert2009l}&Exp.\cite{ParticleDataGroup:2024cfk}\\
		\hline
		$1^1S_0$&$0^{-+}$&$\pi$&$154$&$139.57039(17)$&$\eta$&$538$&$547.862(17)$&$\eta'$&$963$&$957.78(6)$\\
		$1^3S_1$&$1^{--}$&$\rho$&$776$&$775.26(23)$&$\omega$&$782$&$782.66(13)$&$\phi$&$1031$&$1019.460(16)$\\
		$1^3P_0$&$0^{++}$&$a_0(1450)$&$1176$&$1439(34)$&$f_0(1370)$&$1260$&$\!\!\!\{1200..1500\}$&$f_0(1500)$&$1482$&$1522(25)$\\
		$1^3P_1$&$1^{++}$&$a_1(1260)$&$1254$&$1230(40)$&$f_1(1285)$&$1281$&$1281.8(5)$&$f_1(1420)$&$1440$&$1428.4(^{1.5}_{1.3})$\\
		$1^3P_2$&$2^{++}$&$a_2(1320)$&$1317$&$1318.2(6)$&$f_2(1270)$&$1322$&$1275.4(8)$&$f_2^\prime(1525)$&$1524$&$1517.3(2.4)$\\
		$1^1P_1$&$1^{+-}$&$b_1(1235)$&$1258$&$1229.5(3.2)$&$h_1(1170)$&$1258$&$1166(6)$&$h_1(1415)$&$1484$&$1409(^9_8)$\\
		$2^1S_0$&$0^{-+}$&$\pi(1300)$&$1292$&$1300(100)$&$\eta(1295)$&$1294$&$1294(4)$&$\eta(1475)$&$1536$&$1476(4)$\\
		$2^3S_1$&$1^{--}$&$\rho(1450)$&$1486$&$1465(25)$&$\omega(1420)$&$1486$&$1410(60)$&$\phi(1680)$&$1698$&$1680(20)$\\
		$2^3P_0$&$0^{++}$&$a_0(1710)$&$1679$&$1713(19)$&$f_0(1770)$&$1803$&$1804(16)$&$X(1855)$&$1855$&$1857(5)$\\
		$2^3P_1$&$1^{++}$&$a_1(1640)$&$1742$&$1655(16)$&$f_1(1740)$&$1742$&&$f_1(1970)$&$2016$&$1977(15)$\\
		$2^3P_2$&$2^{++}$&$a_2(1700)$&$1779$&$1706(14)$&$f_2(1750)$&$1779$&$1755(10)$&$f_2(2010)$&$2030$&$2010(^{68}_{80})$\\
		$2^1P_1$&$1^{+-}$&$b_1(1720)$&$1721$&&$X(1650)$&$1721$&$1657(7)$&$h_1(1965)$&$2024$&$1965(45)$\\
	$3^1S_0$&$0^{-+}$&$\pi(1800)$&$1788$&$1810(^{9}_{11})$&$\eta(1760)$&$1788$&$1751(15)$&$\eta(2100)$&$2085$&$2050(^{80}_{35})$\\
		$3^3S_1$&$1^{--}$&$\rho(1900)$&$1921$&$1900(20)$&$\omega(1960)$&$1921$&$1960(25)$&$\phi(2170)$&$2119$&$2164(15)$\\	
		
		\end{tabular}
		\end{ruledtabular}
\end{table}		

In the isoscalar sector ($I=0$), it is important to take into account the possible mixing of the $q\bar q\equiv (u\bar u+d\bar d)/\sqrt2$ and $s\bar s$ states. Admixtures of the glueballs is also possible \cite{jrp}.

For the ground ($1^1S_0$) isosinglet ($I=0$) pseudoscalar mesons, we use the mixing scheme, which takes into account the possible glueball component in the $\eta'$ meson \cite{fkr,drb}
 \begin{eqnarray}\label{mixe}
\nonumber \eta&=&\eta\!\left(\frac{u\bar u+d\bar d}{\sqrt2}\right)\cos\phi_P-\eta(s\bar s)\sin\phi_P,\\
 \eta'&=&\eta\!\left(\frac{u\bar u+d\bar d}{\sqrt2}\right)\sin\phi_P\cos\phi_G+\eta(s\bar s)\cos\phi_P\cos\phi_G+G\sin\phi_G,
 \end{eqnarray}
with $G$ denoting the glueball.  The mixing angles are taken to be $\phi_P=(41.6^{+1.0}_{-1.2})^\circ$ and $\phi_G=(28.1^{+3.9}_{-4.0})^\circ$ from the very recent LHCb analysis of the branching fraction ratios between the decays $B^0_{(s)}\to J/\psi\eta^{(')}$ \cite{LHCB2025}.  For the first radial excitation ($2^1S_0$) glueball contributions are neglected ($\phi_G=0$) and the mixing angle is taken to be $\phi_P=1^\circ$  from fitting the $\eta(1295)$ mass, while higher excitations are considered to be pure states ($\phi_G=0$,$\phi_P=0$). In such a scheme the obtained masses of the ground state $\eta$ and $\eta'$ mesons are in better agreement with data than in the mixing scheme without glueball admixture \cite{Ebert2009l}, while $\eta(1295)$, $\eta(1475)$ correspond to the first radial ($2S$) excitation and $\eta(1760)$, $\eta(2100)$ to the second one ($3S$).

The $\omega-\phi$ mixing is also discussed in the literature \citep{gr,kamu,kloe}. It is given by
\begin{eqnarray}\label{mixo}
\nonumber \omega&=&\omega\!\left(\frac{u\bar u+d\bar d}{\sqrt2}\right)\cos\theta_V-\phi(s\bar s)\sin\theta_V,\\
 \phi&=&\omega\!\left(\frac{u\bar u+d\bar d}{\sqrt2}\right)\sin\theta_V+\phi(s\bar s)\cos\theta_V,
 \end{eqnarray}
with the mixing angle $\theta_V=(3.32\pm0.09)^\circ$  experimentally determined by KLOE Collaboration \cite{kloe}. This angle is very small and, thus,  contribution of $q\bar q$ component to $\phi$ meson is usually neglected. We also neglect $q\bar q$ and $s\bar s$ mixing in the radially excited $\omega$ and $\phi$ mesons.

In the orbitally excited sector we see that all calculated scalar ($1^3P_0$) meson masses have values larger than 1~GeV. Scalars with masses less than 1~GeV are considered to be tetraquarks in RQM \cite{Ebert:2008id}. They consist from the scalar diquark and scalar antidiquark. The predicted masses of the lightest scalar tetraquark states are in agreement with experimental data for $f_0(500)$, $K_0^*(700)$, $a_0(980)$ and $f_0(980)$ masses.  Note that the mass of the observed scalar isotriplet ($I=1$) $a_0(1450)$ meson also better agrees with the tetraquark picture with $(qs)(\bar q\bar s)$ composition in which the ground scalar state consisting of an axial-vector diaquark and axial-vector antidiquark has the mass  \cite{Ebert:2008id} in agreement with the experimental value \cite{ParticleDataGroup:2024cfk}, while in $q\bar q$ scenario ($q=u,d$) the predicted mass \cite{Ebert2009l} is more than $6\sigma$ lower than experimental value. Nevertheless, in this paper, we consider  $a_0(1450)$ to be the $1^3P_0$ meson. The RQM prediction for the $2^3P_0$ isotriplet state mass is consistent with the $a_0(1710)$ mass.

Theoretical interpretation of the mixing of the scalar $f_0(1370)$ and $f_0(1500)$ mesons, which are candidates for the $1^3P_0$ states, is uncertain at present. Usually, an additional scalar  $f_0(1710)$ meson is included into analysis. In such a scheme one of these states should be predominantly a glueball \cite{Cheng:2006hu,Cheng:2015iaa,Petrov:2022ipv}. The analysis of Refs.~\cite{Cheng:2006hu,Cheng:2015iaa} indicates that $f_0(1710)$ has the largest glueball component, while $f_0(1500)$ is predominantly flavor octet and the mixing of these states is given by
 \begin{eqnarray}\label{eq:f0}
\begin{pmatrix} f_0(1370) \\ f_0(1500) \\
f_0(1710)\end{pmatrix}=\begin{pmatrix} 0.78(2) & 0.52(3) & -0.36(1)
\\ -0.55(3) & 0.84(2) & 0.03(2) \\  0.31(1) & 0.17(1) & 0.934(4)  \end{pmatrix}
\begin{pmatrix}f_{0}\!\left(\frac{u\bar u+d\bar d}{\sqrt2}\right) \\ f_{0}(s\bar s) \\G \end{pmatrix},
\end{eqnarray}
where $G$ denotes a glueball. { Such a mixing scheme implies that these three mesons contain the full glueball. In our following calculations  the corresponding results are marked with the superscript $^\ast$.} Then we further assume that $f_0(1770)$ belongs to $2^3P_0$ isoscalar state. To get the correct mass of this state the following mixing is introduced
\begin{eqnarray} \label{f02}
 f_0(1770)&=&f_0\!\left(\frac{u\bar u+d\bar d}{\sqrt2}\right)\cos\phi_{f_0}+f_0(s\bar s)\sin\phi_{f_0},\cr
 f_0(1855)&=&-f_0\!\left(\frac{u\bar u+d\bar d}{\sqrt2}\right)\sin\phi_{f_0}+f_0(s\bar s)\cos\phi_{f_0},
 \end{eqnarray}
with the mixing angle $\phi_{f_0}=39.9^\circ$. Then for the partner state we get the mass 1855~MeV, which is consistent with the mass of $X(1855)$ state listed in the Further States of PDG \cite{ParticleDataGroup:2024cfk}. { The results for this mixing are marked by the superscript $^\ddag$.}

{ Recently another mixing scheme for the scalar mesons has been proposed in Refs.~\cite{Klempt1,Klempt2,Klempt3}, which includes the fragmented glueball. In such a scheme, the glueball contribution is distributed over several resonances. The wave function of a scalar meson is taken in the form \cite{Klempt2}
\begin{eqnarray}\label{eq:f0mix}
f_0^{nH}(xxx)&=&\left[f_0\!\left(\frac{u\bar u+d\bar d}{\sqrt2}\right)\cos\phi_n^s-f_0(s\bar s)\sin\phi_n^s\right]\cos\phi_{nH}^G+G\sin\phi_{nH}^G,\nonumber\\
f_0^{nL}(xxx)&=&\left[f_0\!\left(\frac{u\bar u+d\bar d}{\sqrt2}\right)\sin\phi_n^s+f_0(s\bar s)\cos\phi_n^s\right]\cos\phi_{nL}^G+G\sin\phi_{nL}^G,
\end{eqnarray}
$\phi_n^s$ is the scalar mixing angle, and $\phi_{nH}^G$ and $\phi_{nL}^G$ are the meson-glueball mixing angles of the high-mass state $H$ and of the low-mass state
$L$ in the $n$th nonet. The $f_0(1370)$ and $f_0(1500)$ are attributed to the first nonet with $\phi_1^s=(64\pm12)^\circ$, $\phi_{1H}^G=(5\pm 8)^\circ$, $\phi_{1L}^G=(10\pm6)^\circ$, while $f_0(1710)$ and $f_0(1770)$ belong to the second one with $\phi_2^s=(41\pm4)^\circ$, $\phi_{2H}^G=-(29\pm 6)^\circ$, $\phi_{2L}^G=-(20\pm5)^\circ$. The glueball contribution is further distributed in higher excitions of $f_0$ \cite{Klempt2}. We mark the results obtained in this mixing scheme by the superscript $^\dag$.  

 }

In the axial-vector sector we find that the predicted masses of the isotriplet ($I=1$) for $1^3P_1$ and $1^1P_1$ states agree well with the experimental masses of $a_1(1260)$ and $b_1(1235)$. The $a_1(1640)$ meson can be considered as $2^3P_1$ state, while there is no experimental candidate for the $2^1P_1$ state; thus, we use the mass predicted by RQM in the following calculations and denote it as  $b_1(1720)$.

The mixing of the axial-vector isosinglet ($I=0$) meson $1P$ states is in detail discussed in Refs.~\cite{Cheng:2011pb,Cheng:2013cwa} and is given by

 a) for $f_1(1285)$ and $f_1(1420)$
 \begin{eqnarray}\label{f1}
 f_1(1285)&=&f_1\!\left(\frac{u\bar u+d\bar d}{\sqrt2}\right)\sin\theta_{f_1}+f_1(s\bar s)\cos\theta_{f_1},\cr
 f_1(1420)&=&f_1\!\left(\frac{u\bar u+d\bar d}{\sqrt2}\right)\cos\theta_{f_1}-f_1(s\bar s)\sin\theta_{f_1},
 \end{eqnarray}

b) for $h_1(1170)$ and $h_1(1380)$
\begin{eqnarray}\label{h1}
 h_1(1170)&=&h_1\!\left(\frac{u\bar u+d\bar d}{\sqrt2}\right)\sin\theta_{h_1}+h_1(s\bar s)\cos\theta_{h_1},\cr
 h_1(1415)&=&h_1\!\left(\frac{u\bar u+d\bar d}{\sqrt2}\right)\cos\theta_{h_1}-h_1(s\bar s)\sin\theta_{h_1},
 \end{eqnarray}
with  $\theta_{f_1}=69.7^{\circ}$ and $\theta_{h_1}=86.7^{\circ}$, respectively. The calculated and experimental masses of these states are found to be in agreement.
Further, we assume that the $2P$ axial-vector isoscalar states do not mix. In the $q\bar q$ and $s\bar s$ sectors there are no reliable experimental data in the predicted mass region for $f_1(2^3P_1)$ and $h_1(2^1P_1)$ states. The possible candidate for the $q\bar q$
$h_1(2^1P_1)$ state can be the $X(1650)$ meson with $I^G(J^{PC})=0^-(?^{?-})$  from the Further States of PDG \cite{ParticleDataGroup:2024cfk} which has the correct isospin and $C$-parity. In the following calculations we use the predicted mass of the $2^3P_1$ isoscalar state and denote it $f_1(1740)$, since there is no experimental candidate for such state in this mass region. The calculated axial-vector $s\bar s$ $2P$ state masses correspond to $f_1(1970)$ and $h_1(1965)$ mesons also from the Further States of PDG \cite{ParticleDataGroup:2024cfk}.

 In the tensor sector RQM predictions for the masses of the isovector $1^3P_2$ and $2^3P_2$ states agree with experimental values of $a_2(1320)$ and $a_2(1700)$ mesons. Theoretical analysis of the mixing of the isoscalar $1^3P_2$ tensor $f_2(1270)$ and $f'_2(1525)$ states indicate that these states are almost pure $q\bar q$ and $s\bar s$ states. Indeed, the detailed study of such mixing \cite{Cheng:2011fk,Li:2000zb,Li:2018lbd} results in the following structure of these tensor states
 \begin{eqnarray} \label{f2}
 f_2(1270)&=&f_2\!\left(\frac{u\bar u+d\bar d}{\sqrt2}\right)\cos\theta_{f_2}+f_2(s\bar s)\sin\theta_{f_2},\cr
 f'_2(1525)&=&f_2\!\left(\frac{u\bar u+d\bar d}{\sqrt2}\right)\sin\theta_{f_2}-f_2(s\bar s)\cos\theta_{f_2},
 \end{eqnarray}
 with the mixing angle $\theta_{f_2}$ about $(9\pm 1)^{\circ}$ \cite{ParticleDataGroup:2024cfk}. For the $2^3P_2$ isoscalar states we neglect the mixing. The predicted masses are consistent with the experimental values for $f_2(1750)$ and $f_2(2010)$ mesons. Note that the $f_2(1810)$ meson is interpreted as tensor $1^3F_2$ states in RQM  \cite{Ebert2009}.

In the following we use the experimental values of these meson masses for calculating the semileptonic decays.

 \section{Weak $B$ meson decay matrix elements} \label{sec:weak}

{ In this paper we consider the semileptonic  $B$ meson  decays governed by the weak $b\to u$ charged current $J^W_\mu=\overline{u}\gamma_\mu(1-\gamma_5)b$.} Thus, it is necessary to calculate the hadronic matrix element of the corresponding local current between the initial $B$ meson and the final ground state or excited light meson $F$. In RQM such a matrix element is given by \cite{Ebert2003a}
\begin{equation}\label{Matrix}
\langle{F(p_F)|J^W_\mu|B(p_B)}\rangle = \int \dfrac{d^3p d^3q}{(2\pi)^6} \overline{\Psi}_{F\textbf{p}_F}(\textbf{p})\Gamma_\mu(\textbf{p},\textbf{q})\Psi_B\textbf{p}_B(\textbf{q}),
\end{equation}
where $p_M$ ($M=B,F$) is the meson momentum and $\Psi_{M\textbf{p}_M}$ is the meson wave functions projected on the positive energy states and boosted to the moving reference frame with the three-momentum $\textbf{p}_M$. The vertex function $\Gamma$ contains two contributions. The first one is the leading-order vertex function corresponding to the impulse approximation. The second one originates from the interaction between the active and spectator quarks and contains the negative-energy part of the active quark propagator. This contribution is the consequence of the projection on the positive energy states in the quasipotential approach. Explicit expressions for the vertex functions are given in Ref.~\cite{FGK}.

It is convenient to carry out calculations in the rest frame of the decaying  $B$ meson, where the $B$ meson momentum ${\bf p}_B = 0$. Then the final meson $F$ is moving with the recoil momentum ${\bf\Delta} ={\bf p}_F$ and its wave function should be boosted to the moving reference frame. The wave function of the moving meson $\Psi_{F\,{\bf\Delta} }$ is connected with the wave function in the rest frame $\Psi_{F\,{\bf 0}}$ by the transformation \cite{Ebert2003a}
\begin{equation}
\label{wig}
\Psi_{F\,{\bf\Delta}}({\bf
	p})=D_q^{1/2}(R_{L_{\bf\Delta}}^W)D_{\bar q}^{1/2}(R_{L_{
		\bf\Delta}}^W)\Psi_{F\,{\bf 0}}({\bf p}),
\end{equation}
where $R^W$ is the Wigner rotation, $L_{\bf\Delta}$ is the Lorentz boost
from the meson rest frame to a moving one, and  $D^{1/2}_q(R)$ is
the rotation matrix  in spinor representation.

\section{Form factors of the semileptonic $B$ meson decays}	\label{sec:ff}
	
The matrix element $\mathcal{M}$ of the semileptonic decay between meson states is the product of the leptonic part $L_\mu=\overline{\nu}_l\gamma_\mu(1-\gamma_5)l$ and the matrix element of the hadronic current $H^\mu=\langle{F}| \overline{q}\gamma_\mu(1-\gamma_5)b|{B}\rangle$  	
	\begin{equation}
	\mathcal{M}(B\to Fl\nu_l)=\dfrac{G_F}{\sqrt{2}}V_{ub}H^\mu L_\mu,
	\end{equation}
where  $V_{ub}$ is the CKM matrix element and $G_F$ is the Fermi constant. The leptonic part is easily calculated using the lepton spinors and has a simple structure. The hadronic part is significantly more complicated.  It is parameterized by the set of the invariant form factors, which then should be determined  using  nonperturbative approaches within QCD. In this paper we apply RQM for their calculation.  	

\subsection{$B$ meson transitions to the ground and radially excited light mesons }

The hadronic matrix element of the weak current $J^W$ between $B$ meson and ground and radially excited light meson states is parameterized by the following set of the invariant form factors \cite{Ebert2003a}.

\begin{itemize}	
\item[(a)] For $B$ transitions to pseudoscalar $P$ mesons	
\begin{eqnarray}
	\langle P(p_F)|\overline{u}\gamma^\mu b|B(p_B)\rangle &=& f_+(q^2)\left[p^\mu_{B}+p^\mu_F-\dfrac{M_B^2-M_P^2}{q^2}q^\mu\right]+
	f_0(q^2)\dfrac{M_B^2-M_P^2}{q^2}q^\mu,\ \ \ \ \ \ \ \\
	\nonumber \langle P(p_F)|\overline{u}\gamma^\mu \gamma_5b|B(p_B)\rangle & =& 0,
	\end{eqnarray}
	
\item[(b)] For $B$ transitions to vector $V$ mesons	
	\begin{eqnarray}
	\nonumber\langle V(p_F)|\overline{u}\gamma^\mu b|B(p_B)\rangle & =&
		\displaystyle{\frac{2iV(q^2)}{M_B+M_V}}{\epsilon}^{\mu \nu \rho \sigma}{\epsilon}^*_\nu p_{B\rho}p_{F\sigma},\\
		\nonumber\langle V(p_F)|\overline{u}\gamma^\mu \gamma_5b|B(p_B)\rangle &=& 2M_VA_0(q^2)\displaystyle\frac{\epsilon^*\!\! \cdot q}{q^2}q^\mu
		+(M_B+M_V)A_1(q^2)\Big(\epsilon^{*\mu}-\displaystyle\frac{\epsilon^*\!\! \cdot q}{q^2}q^\mu\Big)\\&&\!\!\!\!\! -A_2(q^2)\displaystyle\frac{\epsilon^*\!\! \cdot q}{M_B+M_V}\left[p^\mu_B+p^\mu_F-\dfrac{M^2_B-M^2_V}{q^2}q^\mu\right].\qquad
	\end{eqnarray}
\end{itemize}	
Here $\epsilon^{\mu}$ is the polarization of the vector meson, $q=	p_B-p_F$, and  the following relations among form factors are satisfied at the maximum recoil~($q^2=0$) in order to cancel singularities
	\[f_+(0)=f_0(0),\] \[  A_0(0)=\frac{M_B+M_V}{2M_V}A_1(0)-\frac{M_B-M_V}{2M_V}A_2(0).\]	
	
These form factors are calculated in RQM using the quasipotential approach briefly described in the previous section.	 They are expressed as the overlap integrals of initial and final meson wave functions. The resulting expressions for the form factors take into account all relativistic effects including contributions of the intermediate negative-energy states and relativistic transformations of the wave functions from rest to the moving reference frame.  They are rather cumbersome and are given in Ref. \cite{Ebert2003a}. Let us point out that such relativistic approach allows us to determine the form factor dependence on the transferred momentum squared $q^2$ in the whole accessible kinematical range without additional approximations and extrapolations. The wave functions are numerically known from the meson mass calculations and do not require additional model assumptions, such as Gaussian ones. They include the mixing of the states discussed in Sec.~\ref{sec:mass}.  Note that in the presented analysis we improved our previous results for $B$ meson semileptonic decays to $\eta$, $\eta'$ and $\omega$ \cite{Faustov2022} by more accurate treatment of the mixing (\ref{mixe}) and (\ref{mixo})  of the wave functions and taking the experimental values for the light meson masses instead of theoretical ones in expressions for the form factors.
We find that the numerical results for these decay form factors can be approximated with high accuracy by the following expressions.
\begin{itemize}
\item  For the form factors $f_+(q^2),V(q^2),A_0(q^2)$
\begin{equation}
  \label{fitfv}
  F(q^2)=\frac{F(0)}{\displaystyle\left(1-\frac{q^2}{ M_1^2}\right)
    \left(1-\sigma_1
      \frac{q^2}{M_{B^*}^2}+ \sigma_2\frac{q^4}{M_{B^*}^4}\right)},
\end{equation}

\item For the form factors $f_0(q^2), A_1(q^2),A_2(q^2)$
\begin{equation}
  \label{fita12}
  F(q^2)=\frac{F(0)}{\displaystyle \left(1-\sigma_1
      \frac{q^2}{M_{B^*}^2}+ \sigma_2\frac{q^4}{M_{B^*}^4}\right)},
\end{equation}
\end{itemize}
where the masses $M_{B^*}=5.325$~GeV and $M_1=M_{B^*}$ for $f_+(q^2),V(q^2)$ and $M_1=M_{B}=5.280$~GeV for $A_0(q^2)$. Here $\sigma_{1,2}$ are dimensionless fitted parameters. Their values as well as form factors $F(0)$ and $F(q^2_{\rm max})$ for ground state and their radial excitations are given in Tables~\ref{FF1}, \ref{FF2}.

\begin{table}
	\caption{Form factors of the weak $B$ meson transitions to the ground state  light mesons.}
	\begin{ruledtabular}
		\begin{tabular}{cccccc}
			\text{Decay}&\text{Form factors}&$F(0)$&$F(q^2_{\rm max})$&$\sigma_1$&$\sigma_2$\\
			\hline	
			$B\rightarrow \pi$&$f_+$&0.217&10.96&0.378&$-0.410$\\
			&$f_0$&0.217&1.325&$-0.510$&$-1.500$\\
			$B\rightarrow \rho$&$V$&0.295&2.803&0.8756&0\\
			&$A_0$&0.231&2.132&0.796&$-0.055$\\
			&$A_1$&0.269&0.439&0.540&0\\
			&$A_2$&0.282&1.920&1.340&0.210\\
			$B\rightarrow \eta$&$f_+$&0.170&1.940&0.488&$-0.318$\\
			&$f_0$&0.170&0.511&0.711&$-0.169$\\
			$B\rightarrow \eta'$&$f_+$&0.189&0.777&0.567&$0.197$\\
			&$f_0$&0.189&0.380&1.083&$0.482$\\
			$B\rightarrow \omega$&$V$&0.260&2.6881&0.8770&$-0.072$\\
			&$A_0$&0.202&2.242&0.560&$-0.535$\\
			&$A_1$&0.247&0.4060&0.546&$-0.0036$\\
			&$A_2$&0.263&1.831&1.345&0.202\\
		\end{tabular}\label{FF1}
	\end{ruledtabular}
	
	\end{table}
	\begin{table}
	\caption{Form factors of the weak $B$ meson transitions to the radially excited ($2S$ and $3S$) light mesons.}
	\begin{ruledtabular}
		\begin{tabular}{cccccc}
			\text{Decay}&\text{Form factors}&$F(0)$&$F(q^2_{\rm max})$&$\sigma_1$&$\sigma_2$\\
			\hline
			$B\rightarrow \pi(1300)$&$f_+$&$-0.153$&$-1.701$&2.347&$1.651$\\
			&$f_0$&$-0.153$&$-0.370$&$2.503$&$2.604$\\
			$B\rightarrow \rho(1450)$&$V$&$-0.638$&$-2.626$&0.641&$-0.652$\\
			&$A_0$&$-0.294$&$-1.834$&0.956&$-0.658$\\
			&$A_1$&$-0.202$&$-0.410$&1.803&1.587\\
			&$A_2$&$-0.131$&$-0.270$&3.018&3.922\\
			$B\rightarrow \eta(1295)$&$f_+$&$-0.109$&$-1.222$&2.334&$1.624$\\
			&$f_0$&$-0.109$&$-0.290$&2.425&$2.338$\\
			$B\rightarrow \eta(1475)$&$f_+$&$-0.0027$&$-0.0233$&1.589&$0.189$\\
			&$f_0$&$-0.0027$&$-0.0044$&1.793&$2.053$\\
			$B\rightarrow \omega(1420)$&$V$&$-0.424$&$-1.872$&0.708&$-0.521$\\
			&$A_0$&$-0.132$&$-0.911$&0.964&$-0.634$\\
			&$A_1$&$-0.104$&$-0.295$0&2.405&$-2.233$\\
			&$A_2$&$-0.083$&$-0.190$&3.134&3.912\\
			$B\rightarrow \pi(1800)$&$f_+$&0.0833&0.616&2.537&$1.720$\\
			&$f_0$&0.0833&0.066&$4.106$&$11.166$\\
			$B\rightarrow \rho(1900)$&$V$&0.113&0.957&2.277&0.684\\
			&$A_0$&0.0655&0.448&1.223&$-1.638$\\
			&$A_1$&0.104&0.0867&2.5720&7.637\\
			&$A_2$&0.147&0.0870&3.147&12.153\\
			$B\rightarrow \eta(1760)$&$f_+$&0.0575&0.408&0.197&$-3.431$\\
			&$f_0$&0.0575&0.0356&2.523&$8.913$\\
			$B\rightarrow \omega(1960)$&$V$&0.290&0.822&1.730&$1.652$\\
			&$A_0$&0.0169&0.785&2.227&$-0.653$\\
			&$A_1$&0.0645&0.0337&3.358&$14.670$\\
			&$A_2$&0.1213&0.545&2.024&0.0552\\
		\end{tabular}\label{FF2}
	\end{ruledtabular}
	
	\end{table}

\subsection{$B$ meson transitions to the orbitally excited light mesons }

The hadronic matrix elements of the weak current $J^W$ between $B$ meson and orbitally excited light meson states are parameterized by the following set of the invariant form factors \cite{Ebert2010}.	
\begin{itemize}	
\item[(a)] For $B$ transitions to scalar $S$ mesons		
	\begin{eqnarray}
	\nonumber\langle S(p_F)|\overline{u}\gamma^\mu b|B(p_B)\rangle &=& 0, \\
	 \langle S(p_F)|\overline{u}\gamma^\mu \gamma_5b|B(p_B)\rangle &=&f_+(q^2)(p^\mu_B+p^\mu_F)+f_-(q^2)(p^\mu_B-p^\mu_F).
	\end{eqnarray}
\item[(b)] For $B$ transitions to axial-vector $A$ mesons		
	\begin{eqnarray}
	\nonumber\langle A(p_F)|\overline{u}\gamma^\mu b|B(p_B)\rangle &=& (M_B+M_A)V_1(q^2){\epsilon}^{*\mu} +[V_2(q^2)p^\mu_B+V_3(q^2)p^\mu_F]\displaystyle \frac{{\epsilon}^*\!\!\cdot q}{M_B}, \nonumber\\
	\langle A(p_F)|\overline{u}\gamma^\mu \gamma_5b|B(p_B)\rangle& =& \displaystyle\frac{2iA(q^2)}{M_B+M_A}{\epsilon}^{\mu \nu \rho \sigma}{\epsilon}^*_\nu p_{B\rho}p_{F\sigma}.
	\end{eqnarray}
\item[(c)] For $B$ transitions to tensor $T$ mesons		
	\begin{eqnarray}
	\langle T(p_F)|\overline{u}\gamma^\mu b|B(p_B)\rangle &=& \displaystyle\frac{2iV(q^2)}{M_B+M_T}{\epsilon}^{\mu \nu \rho \sigma}{\epsilon}^*_{\nu \alpha}\displaystyle\frac{p^\alpha_B}{M_B}p_{B\rho}p_{F\sigma},\\
\nonumber	\langle T(p_F)|\overline{u}\gamma^\mu \gamma_5b|B(p_B)\rangle &=& (M_B+M_T)A_1(q^2){\epsilon}^{*\mu}_{\alpha}\displaystyle\frac{p^\alpha_B}{M_B}+[A_2(q^2)p^\mu_B+A_3(q^2)p^\mu_F]{\epsilon}^*_{\alpha\beta}\displaystyle\frac{p^\alpha_Bp^\beta_B}{M^2_B}.
	\end{eqnarray}
	\end{itemize}
Here 	${\epsilon}_{\alpha\beta}$ is the polarization of the tensor meson.

These form factors are calculated in RQM. Explicit expressions for the form factors in terms of the overlap integrals of the meson wave functions with the account of all relativistic contributions are given in Ref.~ \cite{Ebert2010}. Again the numerical results for these decay form factors can be approximated with high accuracy by the following expressions:
		\begin{equation}
	\label{Ap1}
	F(q^2)=\dfrac{F(0)}{\Big(1-\sigma_1\dfrac{q^2}{M_{B^*}^2}+\sigma_2\dfrac{q^4}{M_{B^*}^4}+\sigma_3\dfrac{q^6}{M_{B^*}^6}\Big)},
	\end{equation}
	where $\sigma_{1,2,3}$ are dimensionless fitted parameters and the mass of the vector $B^*$ meson $M_{B^*}=5.325$~GeV  was used for normalization. The values of form factors $F(0), F(q_{\rm max}^2)$ and parameters $\sigma_{1,2,3}$  fitted to numerically calculated form factors in the whole $q^2$ range are given in Tables~\ref{FF3}--\ref{FF6}. The values of the $q_{\rm max}^2$ were evaluated for the central values of the experimental meson masses given in Table~\ref{Mass1}.

\begin{table}
	\caption{Form factors of the weak $B$ meson transitions to the orbitally excited ($1P$) light mesons with isospin $I=1$.}
	\begin{ruledtabular}
		\begin{tabular}{ccccccc}
			\text{Decay}&\text{Form factors}&$F(0)$&$F(q^2_{\rm max})$&$\sigma_1$&$\sigma_2$&$\sigma_3$\\
			\hline
			$B\rightarrow a_0(1450)$ & $f_+$ & 0.309 & 0.859 & 1.586 & $-0.516$& 2.302\\
			& $f_-$& $-0.641$ & $-1.688$& 1.834 & 2.138&$-1.741$\\
			$B\rightarrow a_1(1260)$& $A$ &$-0.412$ &$-1.485$&2.045&1.224&0.250\\
			&$V_1$&$-0.124$&0.197&0.700&$-0.372$&$-1.073$\\
			&$V_2$&0.0815&$-0.221$&1.622&$-0.335$&1.334\\
			&$V_3$&0.593&2.562&1.004&0.100&$-1.220$\\
			$B\rightarrow b_1(1235)$& $A$ &$-0.0339$ &0.0669&0.401&6.283&$-13.529$\\
			&$V_1$&$-0.0319$&$-0.0590$&$-9.248$&$-3.385$&1.012\\
			&$V_2$&$-0.106$&$-0.292$&2.827&4.064&$-1.872$\\
			&$V_3$&$-0.375$&$-1.058$&1.360&$-0.105$&0.908\\
			$B\rightarrow a_2(1320)$& $V$ &$-0.241$ &$-0.968$&2.662&3.850&$-4.979$\\
			&$A_1$&$-0.172$&$-0.294$&1.409&1.795&$-2.296$\\
			&$A_2$&$-0.0089$&$-0.365$&6.269&16.310&$-17.010$\\
			&$A_3$&$-0.0094$&$-0.011$&1.907&0.363&9.001\\
			\end{tabular}\label{FF3}
	\end{ruledtabular}
\end{table}
		\begin{table}
	\caption{Form factors of the weak $B$ meson transitions to the orbitally and radially excited ($2P$) light mesons with isospin $I=1$.}
	\begin{ruledtabular}
		\begin{tabular}{ccccccc}
			\text{Decay}&\text{Form factors}&$F(0)$&$F(q^2_{\rm max})$&$\sigma_1$&$\sigma_2$&$\sigma_3$\\
			\hline	
			$B\rightarrow a_0(1710)$ & $f_+$ & 0.168 & 0.402 & 2.636 & $4.428$& $-3.209$\\
			& $f_-$& $-0.330$ & $-0.629$& 1.814 & 1.402&$0.631$\\
			$B\rightarrow a_1(1640)$& $A$ &$0.379$ &$2.553$&5.058&15.941&$-19.404$\\
			&$V_1$&$0.0978$&0.0371&1.763&$-21.816$&$55.334$\\
			&$V_2$&0.0986&$1.915$&2.358&$4.185$&$-7.585$\\
			&$V_3$&$-0.424$&$-10.968$&5.416&17.171&$-21.498$\\
			$B\rightarrow b_1(1720)$& $A$ &$0.0025$ &0.201&7.653&20.483&$-18.581$\\
			&$V_1$&$-0.0474$&$-0.243$&$5.639$&$18.218$&$-21.554$\\
			&$V_2$&$0.0473$&$0.0408$&7.436&21.787&$-9.695$\\
			&$V_3$&$0.394$&$3.809$&5.225&$16.009$&$-19.715$\\
			$B\rightarrow a_2(1700)$& $V$ &$0.320$ &$1.636$&4.568&14.046&$-23.967$\\
			&$A_1$&$0.202$&$0.234$&2.450&1.387&$16.138$\\
			&$A_2$&$-0.0079$&$0.363$&8.094&23.894&$-25.542$\\
			&$A_3$&$0.0046$&$0.072$&5.462&26.232&$-39.678$\\
			
		\end{tabular}\label{FF4}
	\end{ruledtabular}
\end{table}
\begin{table}
	\caption{Form factors of the weak $B$ meson transitions to the orbitally excited ($1P$) light mesons with isospin $I=0$. { The superscripts $^\ast$ and $^\dag$ correspond to the mixing schemes (\ref{eq:f0}) and (\ref{eq:f0mix}), respectively.} }
	\begin{ruledtabular}
		\begin{tabular}{ccccccc}
			\text{Decay}&\text{Form factors}&$F(0)$&$F(q^2_{\rm max})$&$\sigma_1$&$\sigma_2$&$\sigma_3$\\
			\hline
			$B\rightarrow f_0(1370)$ & $f_+$ & $\begin{array}{c}
			0.164^\ast\\0.186^\dag
			\end{array}$ & $\begin{array}{c}0.509^\ast\\ 0.578^\dag\end{array}$& 1.605 & $-0.557$& 2.227\\
			& $f_-$& $\begin{array}{c}
			-0.328^\ast\\-0.372^\dag
			\end{array}$ & $\begin{array}{c}
			-0.891^\ast\\-1.01^\dag
			\end{array}$& 1.741 & 1.527&$-0.878$\\
			$B\rightarrow f_0(1500)$ & $f_+$ & $\begin{array}{c}
			-0.127^\ast\\-0.101^\dag
			\end{array}$ &$\begin{array}{c}
			-0.300^\ast\\-0.239^\dag
			\end{array}$ & 1.539 & $-0.630$& 2.816\\
			& $f_-$& $\begin{array}{c}
			0.281^\ast\\0.224^\dag
			\end{array}$ & $\begin{array}{c}
			0.734^\ast\\0.583^\dag
			\end{array}$& 1.992 & 3.052&$-3.084$\\
			$B\rightarrow f_0(1710)$ & $f_+$ & $0.0739^\ast$ & $0.161^\ast$ & 2.314 & $3.835$& $-3.065$\\
			& $f_-$& $-0.204^\ast$ & $-0.489^\ast$& 1.962 & 4.306&$-6.417$\\
			$B\rightarrow f_1(1285)$& $A$ &$-0.204$ &$-0.720$&2.191&1.919&$-0.506$\\
			&$V_1$&$-0.0541$&0.116&0.243&$-2.567$&$1.083$\\
			&$V_2$&0.0333&$-0.145$&2.404&$1.864$&$-0.444$\\
			&$V_3$&0.270&1.380&2.439&4.417&$-4.651$\\
			$B\rightarrow f_1(1420)$& $A$ &$-0.0923$ &$-0.311$&2.523&3.771&$-2.905$\\
			&$V_1$&$-0.0207$&$0.0577$&$0.761$&$-1.615$&0.343\\
			&$V_2$&$0.0083$&$-0.143$&4.812&10.028&$-8.207$\\
			&$V_3$&$0.120$&$0.688$&2.917&$6.971$&$-8.431$\\
			$B\rightarrow h_1(1170)$& $A$ &$-0.0202$ &$0.0656$&$-1.300$&$-1.472$&$-4.994$\\
			&$V_1$&$-0.0219$&$-0.0313$&$-1.592$&$-3.942$&0.717\\
			&$V_2$&$-0.0682$&$-0.186$&2.408&2.134&$0.203$\\
			&$V_3$&$-0.244$&$-0.680$&1.341&$-0.288$&$1.232$\\
			$B\rightarrow h_1(1415)$& $A$ &$-0.0016$ &$-0.0024$&1.276&0.813&$0.740$\\
			&$V_1$&$-0.0004$&$0.0075$&$4.873$&$9.848$&$-7.617$\\
			&$V_2$&$-0.0066$&$-0.0143$&2.714&5.475&$-4.268$\\
			&$V_3$&$-0.0197$&$-0.0834$&2.532&$5.901$&$-7.280$\\
			$B\rightarrow f_2(1270)$& $V$ &$-0.160$ &$-0.597$&2.633&3.543&$-4.069$\\
			&$A_1$&$-0.114$&$-0.191$&1.319&0.963&$-0.590$\\
			&$A_2$&$-0.0344$&$-0.166$&6.964&18.531&$-18.218$\\
			&$A_3$&$-0.0062$&$-0.0079$&1.900&0.0459&8.823\\
		$B\rightarrow f'_2(1525)$& $V$ &$-0.0368$ &$-0.142$&3.149&6.694&$-10.552$\\
			&$A_1$&$-0.0273$&$-0.0518$&2.095&6.618&$-12.701$\\
			&$A_2$&$-0.0038$&$-0.0262$&3.395&3.439&$-1.909$\\
			&$A_3$&$-0.0012$&$-0.014$&2.353&4.690&2.721\\	
			
		\end{tabular}\label{FF5}
	\end{ruledtabular}
\end{table}

\begin{table}
	\caption{Form factors of the weak $B$ meson transitions to the orbitally and radially excited ($2P$) light mesons with isospin $I=0$. { The superscripts $^\ddag$ and $^\dag$ correspond to the mixing schemes (\ref{f02}) and (\ref{eq:f0mix}), respectively.}}
	\begin{ruledtabular}
		\begin{tabular}{ccccccc}
			\text{Decay}&\text{Form factors}&$F(0)$&$F(q^2_{\rm max})$&$\sigma_1$&$\sigma_2$&$\sigma_3$\\
			\hline
	$B\rightarrow f_0(1710)$ & $f_+$ & $-0.139^\dag$ & $-0.399^\dag$ & 2.698 & $5.934$& $-7.132$\\
			& $f_-$& $0.295^\dag$ & $0.673^\dag$& 2.318 & 3.198&$-1.857$\\			
		$B\rightarrow f_0(1770)$ & $f_+$ & $\begin{array}{c}
			-0.173^\ddag\\-0.149^\dag
			\end{array}$ & $\begin{array}{c}
			-0.460^\ddag\\-0.396^\dag
			\end{array}$ & 2.698 & $5.934$& $-7.132$\\
			& $f_-$& $\begin{array}{c}
			0.367^\ddag\\0.316^\dag
			\end{array}$ & $\begin{array}{c}
			0.818^\ddag\\0.704^\dag
			\end{array}$& 2.318 & 3.198&$-1.857$\\
	
			$B\rightarrow f_0(1855)$ & $f_+$ & $-0.148$ &$-0.367$ & 2.679 & $6.260$& $-7.912$\\
			& $f_-$& $0.327$ & $0.686$& 2.168 & 2.788&$-1.472$\\
			$B\rightarrow f_1(1740)$& $A$ &$0.260$ &$1.332$&5.195&16.686&$-20.486$\\
			&$V_1$&$0.00629$&$-0.105$&0.909&$2.360$&$-9.119$\\
			&$V_2$&0.0484&$0.730$&5.417&$14.939$&$-16.891$\\
			&$V_3$&$-0.205$&$-4.342$&6.870&20.990&$-23.360$\\
			$B\rightarrow h_1(1650)$& $A$ &$0.0107$ &$0.141$&$4.550$&$9.120$&$-7.981$\\
			&$V_1$&$-0.0214$&$-0.274$&$6.680$&$19.498$&$-20.220$\\
			&$V_2$&$0.112$&$-0.104$&$-0.139$&2.263&$-12.106$\\
			&$V_3$&$0.284$&$1.887$&4.190&$11.457$&$-13.732$\\
			$B\rightarrow f_2(1750)$& $V$ &$0.204$ &$0.878$&4.692&13.892&$-21.749$\\
			&$A_1$&$0.130$&$0.181$&3.090&4.535&$7.993$\\
			&$A_2$&$-0.0073$&$-0.142$&4.957&4.271&$5.969$\\
			&$A_3$&$0.0028$&$0.0054$&6.618&31.546&$-49.341$\\
			
		\end{tabular}\label{FF6}
	\end{ruledtabular}
\end{table}

	We estimate the uncertainties of the calculated form factors of the semileptonic $B$ decays to the ground and excited states of light mesons, which originate from the model parameters and fitting, to be less than 5\%.

\section{Semileptonic $B$ meson decays}\label{sec:sem}

We use the calculated form factors for the evaluation of the semileptonic decays branching fractions and different asymmetry and polarization observables. This can be most conveniently done in the framework of the helicity formalism. The differential decay rate of the semileptonic $B$ decays can be expressed in the following form~\cite{Ivanov2019}:
\begin{eqnarray}\label{Gamma}
\nonumber\dfrac{d\Gamma(B\rightarrow Fl^+\nu_l)}{dq^2d\cos\theta}&=&\dfrac{G^2_F}{(2\pi)^3}|V_{ub}|^2\dfrac{\lambda^{1/2}(q^2-m_l^2)^2}{64M_B^3q^2}
\Big[(1+\cos^2\theta)\mathcal{H}_U +2\sin^2\theta \mathcal{H}_L +2\cos\theta \mathcal{H}_P\\
&&+2\delta_l\Big(\sin^2\theta \mathcal{H}_U+2\cos^2\theta \mathcal{H}_L+2\mathcal{H}_S-4\cos\theta \mathcal{H}_{SL}\Big) \Big],
\end{eqnarray}	
where $\lambda\equiv\lambda(M_{B}^2,M_F^2,q^2)= M_{B}^4+M_F^4+q^4-2(M_{B}^2M_F^2+M_{B}^2q^2+M^2_Fq^2)$, $m_l$ is
the lepton mass, $\delta_l=\frac{m_l^2}{2q^2}$, and the polar angle $\theta$ is the angle between the momentum of the charged lepton in the rest frame of the intermediate $W$-boson and the direction opposite to the final $F$ meson momentum in the rest frame of $B$. The bilinear combinations $\mathcal{H}_I$ ($I=U,L,P,S,SL$) of the helicity components of the hadronic tensor are defined by
\begin{eqnarray}\label{hc}
&&\mathcal{H}_U=|H_+|^2+|H_-|^2,\qquad \mathcal{H}_L=|H_0|^2,\qquad \mathcal{H}_P=|H_+|^2-|H_-|^2,\cr
&& \mathcal{H}_S=|H_t|^2,\qquad\qquad \qquad\mathcal{H}_{SL}=\Re(H_0H^\dagger_t),
\end{eqnarray}
and the helicity amplitudes are expressed through invariant form factors~\cite{Ivanov2019,Ebert2010,FGK}.
\begin{itemize}

\item For $B\to P$ transitions
\begin{eqnarray} \label{dp}
H_\pm =0, \qquad H_0=\dfrac{\lambda^{1/2}}{\sqrt{q^2}}f_+(q^2),\qquad H_t=\dfrac{1}{\sqrt{q^2}}\Big(M^2_{B}-M^2_P\Big)f_0(q^2).
\end{eqnarray}

\item For $B\to V$ transitions
\begin{eqnarray}\label{dv}
H_\pm&=&\dfrac{\lambda^{1/2}}{M_{B}+M_V}\left[V(q^2)\mp \dfrac{(M_{B}+M_V)^2}{\lambda^{1/2}}A_1(q^2)\right],\cr
H_0&=&\dfrac{1}{2M_V\sqrt{q^2}}\left[(M_{B}+M_V)(M^2_{B}-M_V^2-q^2)A_1(q^2)-\dfrac{\lambda}{M_{B}+M_V}A_2(q^2)\right],\cr
H_t&=&\dfrac{\lambda^{1/2}}{\sqrt{q^2}}A_0(q^2).
\end{eqnarray}

\item For $B\to S$ transitions
\begin{eqnarray}\label{ds}
H_\pm&=&0,\qquad H_0=\dfrac{\lambda^{1/2}}{\sqrt{q^2}}f_+(q^2),\cr H_t&=&\dfrac{1}{\sqrt{q^2}}[(M^2_{B}-M^2_S)f_+(q^2)+q^2f_-(q^2)].
\end{eqnarray}

\item For $B\to A$ transitions
\begin{eqnarray}\label{da}
\nonumber H_\pm&=&(M_{B}+M_{A})V_1(q^2)\pm\dfrac{\lambda^{1/2}}{M_{B}+M_{A}}A(q^2),\\
\nonumber H_0&=&\dfrac{1}{2M_{A}\sqrt{q^2}}\Big\{(M_{B}+M_{A})(M^2_{B}-M^2_{A}-q^2)V_1(q^2)\\
\nonumber &&\qquad\qquad\quad+\dfrac{\lambda}{2M_{B}}[V_2(q^2)+V_3(q^2)] \Big\},\\
\nonumber H_t&=&\dfrac{\lambda^{1/2}}{2M_{A}\sqrt{q^2}}\Bigg\{(M_{B}+M_{A})V_1(q^2)+\dfrac{M_{B}^2-M^2_{A}}{2M_{B}}[V_2(q^2)+V_3(q^2)]\\&&\qquad\qquad\quad+\dfrac{q^2}{2M_{B}}[V_2(q^2)-V_3(q^2)] \Bigg\}.
\end{eqnarray}

\item For $B\to T$ transitions
\begin{eqnarray}\label{dt}
\nonumber H_\pm&=&\dfrac{\lambda^{1/2}}{2\sqrt{2}M_{B}M_T}\left\{(M_{B}+M_T)A_1(q^2)\pm\dfrac{\lambda^{1/2}}{M_{B}+M_T}V(q^2) \right\},\\
\nonumber H_0&=&\dfrac{\lambda^{1/2}}{2\sqrt{6}M_{B}M_{T}^2\sqrt{q^2}}\Big\{(M_{B}+M_{T})(M^2_{B}-M^2_{T}-q^2)A_1(q^2)\\
&&\nonumber\qquad\qquad\qquad\qquad\ \ +\dfrac{\lambda}{2M_{B}}[A_2(q^2)+A_3(q^2)] \Big\},\\
\nonumber H_t&=&\sqrt{\dfrac{2}{3}}\dfrac{\lambda}{4M_BM_{T}^2\sqrt{q^2}}\Bigg\{(M_{B}+M_T)A_1(q^2)\\
&&\qquad\qquad+\dfrac{M_{B}^2-M^2_T}{2M_{B}}[A_2(q^2)+A_3(q^2)]+\frac{q^2}{2M_{B}}[A_2(q^2)-A_3(q^2)] \Bigg\}.\quad
\end{eqnarray}
\end{itemize}
Here the subscripts $\pm,0,t$ denote transverse, longitudinal, and time helicity components, respectively.

The expression for the differential decay rate (\ref{Gamma}) normalized by the decay rate  integrated over $\cos\theta$,
\begin{equation}
  \label{eq:dg}
d\Gamma/dq^2\equiv  \frac{d\Gamma(B\to
  F l^+\nu_l)}{dq^2}=\frac{G_F^2}{(2\pi)^3}
  |V_{ub}|^2\frac{\lambda^{1/2}q^2}{24M_{{B}}^3}\left(1-\frac{m_l^2}{q^2}\right)^2\mathcal{H}_{\rm tot},
\end{equation}
 can be rewritten as
\begin{eqnarray}
\dfrac{1}{d\Gamma/dq^2}\dfrac{d\Gamma(B\to Fl^+\nu_l)}{dq^2d(\cos\theta)}=\dfrac{1}{2}\Big[1-\frac{1}{3}C^l_F(q^2)\Big]+A_{FB}(q^2)\cos\theta+\frac{1}{2}C^l_F(q^2)\cos^2\theta,
\end{eqnarray}
where the total helicity structure
\begin{eqnarray}
\mathcal{H}_{\rm tot}=\mathcal{H}_U+\mathcal{H}_L+\delta_l(\mathcal{H}_U+\mathcal{H}_L+3\mathcal{H}_S).
\end{eqnarray}
The forward-backward asymmetry is defined by
\begin{eqnarray}
\label{Afb}
A_{FB}(q^2)=\frac{\int_{0}^{1} d\cos\theta \frac{d\Gamma}{dq^2d\cos\theta}-\int_{-1}^{0} d\cos\theta \frac{d\Gamma}{dq^2d\cos\theta}}{\int_{0}^{1} d\cos\theta \frac{d\Gamma}{dq^2d\cos\theta}+\int_{-1}^{0} d\cos\theta \frac{d\Gamma}{dq^2d\cos\theta}}=\frac{3}{4}\frac{\mathcal{H}_P-4\delta_l\mathcal{H}_{SL}}{\mathcal{H}_{\rm tot}},
\end{eqnarray}
and the lepton-side convexity parameter, which is the second derivative of the distribution  over $\cos\theta$, is given by
\begin{equation}
\label{Clf}
C^l_F(q^2)=\frac{3}{4}(1-2\delta_l)\frac{\mathcal{H}_U-2\mathcal{H}_L}{\mathcal{H}_{\rm tot}}.
\end{equation}

Other useful observables are the longitudinal polarization of the final charged lepton $l$ defined by
\begin{eqnarray}
\label{Pll}
P^l_L(q^2)=\frac{\mathcal{H}_U+\mathcal{H}_L-\delta_l(\mathcal{H}_U+\mathcal{H}_L+3\mathcal{H}_S)}{\mathcal{H}_{\rm tot}}.
\end{eqnarray}
and its transverse polarization
\begin{eqnarray}
\label{Plt}
P^l_T(q^2)=-\dfrac{3\pi\sqrt{\delta_l}}{4\sqrt{2}}\frac{\mathcal{H}_P+2\mathcal{H}_{SL}}{\mathcal{H}_{\rm tot}}.
\end{eqnarray}

For the semileptonic $B$ decays to the vector $V$ meson, which then decays to two pseudoscalar mesons $V\to P_1P_2$, the differential distribution in the angle $\theta^*$, defined as the polar angle between the vector meson $V$ momentum in the $B$ meson rest frame and the pseudoscalar meson $P_1$ momentum in the rest frame of the vector meson $V$, in the narrow width approximation is given by \cite{Ivanov2019}
\begin{equation}
  \label{eq:mpdr}
\frac1{d\Gamma/dq^2} \frac{d\Gamma(B\to
  V(\to P_1P_2)l^+\nu_l)}{dq^2d(\cos\theta^*)} =
\frac34\left[2F_L(q^2)\cos^2\theta^* +F_T(q^2)\sin^2\theta^*\right].
\end{equation}
Here the longitudinal polarization fraction of the final vector meson has
the form
\begin{equation}
\label{Fl}
F_L(q^2)=\frac{\mathcal{H}_L+\delta_l(\mathcal{H}_L+3\mathcal{H}_S)}{\mathcal{H}_{\rm tot}},
\end{equation}
and its transverse polarization fraction $F_T(q^2)=1- F_L(q^2)$.

\section{Results and discussion}\label{sec:results}

To calculate the semileptonic $B$ decays branching fractions and differential distributions we substitute the corresponding form factors in the expressions for the helicity amplitudes (\ref{dp})--(\ref{dt})  of the hadronic structure (\ref{hc}), which enter the expression (\ref{Gamma}) for the differential decay rate. Note that in our error estimates we include uncertainties of the form factor calculations, uncertainties in the measured light meson masses, uncertainties in the $|V_{ub}|$ matrix element (see discussion below), but we do not include uncertainties in the mixing angles. Such uncertainties are important for the isoscalar mesons, especially, for the orbitally excited ones. However, it is difficult to estimate them.

\subsection{Semileptonic $B$ decays to the light meson ground states  and the $|V_{ub}|$ extraction }

First, we consider semileptonic $B$ decays to the ground states of the light mesons. The experimental data are available for such decays at present. The modes only with light leptons ($l=e$ or $\mu$) were measured. We use these data in order to extract the value of the CKM matrix element $|V_{ub}|$. In Table~\ref{Vub} the RQM predictions for the branching fractions normalized by $|V_{ub}|^2$ are confronted with the averaged experimental values form PDG \cite{ParticleDataGroup:2024cfk}. Note that the very recent Belle II measurement \cite{belle2025} provides the total branching fractions ${\rm Br}(B^0\to\pi^-l^+\nu_l)=(1.516\pm0.042\pm0.059)\times 10^{-4}$ and ${\rm Br}(B^+\to\rho^0l^+\nu_l)=(1.625\pm0.079\pm0.180)\times 10^{-4}$, which are in good agreement with PDG values.  The extracted values of $|V_{ub}|$ for each decay channel are given in the last column and the last line gives the  value averaged over all presented decay channels. Experimental and theoretical errors are added in quadrature. We see that decays to the pseudoscalar mesons favor a slightly larger $|V_{ub}|$  values than decays to the vector mesons. However, they are compatible within error bars.

\begin{table}
	\caption{Extraction of the $|V_{ub}|$ values from the excremental data from the total branching fractions.}
	\begin{ruledtabular}
		\begin{tabular}{cccc}			
			\text{Decay}&Br (RQM)&Br (PDG)($\times10^{-5})$ &$|V_{ub}|(\times10^{-3})$\\
			\hline
$B^0\rightarrow \pi^- l^+\nu_l$&$(8.34\pm 0.90)|V_{ub}|^2$&$15.0\pm0.5$&$4.24\pm0.24$\\
$B^+\rightarrow \pi^0 l^+\nu_l$ &$(4.50\pm0.50)|V_{ub}|^2$&$7.80\pm0.27$&$4.16\pm0.24$\\
$B^0\rightarrow \rho^- l^+\nu_l$&$(20.0\pm 2.1)|V_{ub}|^2$&$29.4\pm2.1$&$3.83\pm0.24$\\
$B^+\rightarrow \rho^0 l^+\nu_l$ &$(10.8\pm 1.1)|V_{ub}|^2$&$15.8\pm1.1$&$3.82\pm0.24$\\
	$B^+\rightarrow \eta l^+\nu_l$&$(2.12\pm0.22)|V_{ub}|^2$&$3.5\pm0.4$&$4.06\pm0.31$\\
$B^+\rightarrow \eta' l^+\nu_l$ &$(1.64\pm0.18)|V_{ub}|^2$&$2.4\pm0.7$&$3.82\pm0.60$\\
$B^+\rightarrow \omega l^+\nu_l$ &$(8.81\pm 0.92)|V_{ub}|^2$&$11.9\pm0.9$&$3.68\pm0.24$\\
\hline
Average&&&$3.95\pm0.15$\\
	
		\end{tabular}\label{Vub}
	\end{ruledtabular}
\end{table}	

The $|V_{ub}|$  values can be also extracted from the experimental data for the partial decay branching fractions, measured in sequential $q^2$ bins, for $B^0\to \pi^-l^+\nu_l$ and decay spectra for $B\to\rho l\nu_l$ and $B\to\omega l\nu_l$. Such measurements have been performed by the Belle and BaBar Collaborations. The averaged data on partial decay rates is given by the Heavy Flavor Averaging Group (HFLAV) \cite{hflav}. We find that RQM results agree with these data.  In Table~\ref{Vub2} we present the $|V_{ub}|$  values extracted from different sources of the averaged partial decay rates \cite{hflav} and very recent Belle II data on $B^0\to\pi^-l^+\nu_l$ and $B^+\to\rho^0 l^+\nu_l$ decays \cite{belle2025}. Again we find that decays to the pseudoscalar mesons favor larger values than decays to the vector mesons. In the last line the averaged value of $|V_{ub}|$ is given. We see that the central $|V_{ub}|$ values extracted from averaged HFLAV and Belle II data coincide. They are slightly larger than the averaged one extracted from the total branching fractions, but these values agree well within errors. Averaging them we finally get
\begin{equation}\label{eq:vub}
|V_{ub}|=(4.00\pm0.11)\times 10^{-3}.
\end{equation}
This value is in a very good agreement with the $|V_{ub}|=(4.06\pm0.12\pm 0.11)\times 10^{-3}$ extracted from the inclusive decays \cite{hflav} and HFLAV averaged value form exclusive decays $|V_{ub}|=(3.75\pm0.06\pm 0.19)\times 10^{-3}$ \cite{hflav}, but with the higher central value. In the following we use the value (\ref{eq:vub}) of $|V_{ub}|$ in our calculations of the branching fractions of the semileptonic $B$ decays to light mesons.

\begin{table}
	\caption{Extraction of the $|V_{ub}|$ values from the excremental data from the averaged partial decay rates.}
	\begin{ruledtabular}
		\begin{tabular}{ccc}			
			\text{Decay} &$|V_{ub}|(\times10^{-3})$ (HFLAV) \cite{hflav}&$|V_{ub}|(\times10^{-3})$ (Belle II)\cite{belle2025}\\
			\hline
$B\rightarrow \pi l\nu_l$&$4.21\pm0.12$&$4.20\pm0.18$\\
$B\rightarrow \rho l\nu_l$&$3.72\pm0.21$&$3.80\pm0.24$\\
$B\rightarrow \omega l\nu_l$ &$3.84\pm0.24$\\
\hline
Average&$4.05\pm0.15$&$4.05\pm0.20$\\
	
		\end{tabular}\label{Vub2}
	\end{ruledtabular}
\end{table}

The calculated branching fractions of the semileptonic $B$ decays to the ground states of light mesons are given in Table~\ref{Brg}. In Table~\ref{Brcompg} we compare the RQM results with the predictions of the covariant light-front quark model (CLFQM) \cite{klzdw} and available experimental data \cite{ParticleDataGroup:2024cfk}. The presented theoretical results agree with each other and experiment.

\begin{table}
	\caption{Branching fractions of the semileptonic $B$ decays to the ground state light mesons ($\times 10^{-5}$).}
	\begin{ruledtabular}
		\begin{tabular}{cccc}			
			\text{Decay}&\text{Br}&\text{Decay}&\text{Br}\\
			\hline
			$B^0\rightarrow \pi^- e^+\nu_e$&$13.4\pm 1.4$&$B^+\rightarrow \pi^0 e^+\nu_e$ &$7.20\pm0.75$\\
			$B^0\rightarrow \pi^- \mu^+\nu_\mu$&$13.3\pm1.4$&$B^+\rightarrow \pi^0 \mu^+\nu_\mu$ &$7.20\pm0.75$\\
	$B^0\rightarrow \pi^- \tau^+\nu_\tau$&$8.43\pm0.87$&$B^+\rightarrow \pi^0 \tau^+\nu_\tau$ &$4.54\pm0.47$\\	
		$B^0\rightarrow \rho^- e^+\nu_e$&$32.2\pm 3.4$&$B^+\rightarrow \rho^0 e^+\nu_e$ &$17.4\pm1.8$\\
			$B^0\rightarrow \rho^- \mu^+\nu_\mu$&$32.1\pm3.4$&$B^+\rightarrow \rho^0 \mu^+\nu_\mu$ &$17.3\pm1.8$\\
	$B^0\rightarrow \rho^- \tau^+\nu_\tau$&$18.0\pm1.9$&$B^+\rightarrow \rho^0 \tau^+\nu_\tau$ &$9.72\pm0.98$\\	
	$B^+\rightarrow \eta e^+\nu_e$&$3.40\pm 0.37$&$B^+\rightarrow \eta' e^+\nu_e$ &$2.63\pm0.28$\\
			$B^+\rightarrow \eta \mu^+\nu_\mu$&$3.39\pm0.37$&$B^+\rightarrow \eta' \mu^+\nu_\mu$ &$2.62\pm0.28$\\
	$B^+\rightarrow \eta \tau^+\nu_\tau$&$2.20\pm0.25$&$B^+\rightarrow \eta' \tau^+\nu_\tau$ &$1.51\pm0.17$\\	
	$B^+\rightarrow \omega e^+\nu_e$ &$14.1\pm1.5$
			&$B^+\rightarrow \omega \mu^+\nu_\mu$ &$14.1\pm1.5$\\
	$B^+\rightarrow \omega \tau^+\nu_\tau$ &$7.91\pm0.80$\\	
	
		\end{tabular}\label{Brg}
	\end{ruledtabular}
\end{table}	

\begin{table}
	\caption{Comparison of the calculated in RQM branching fractions of the semileptonic $B$ decays to the ground state light mesons with theoretical predictions in the covariant light-front quark model (CLFQM)\cite{klzdw} and experimental data \cite{ParticleDataGroup:2024cfk}($\times 10^{-5}$).}
	\begin{ruledtabular}
		\begin{tabular}{@{\!\!}ccc@{\!}cccc@{}c@{\!\!}}			
			\text{Decay}&\text{RQM}& CLFQM\cite{klzdw}& PDG\cite{ParticleDataGroup:2024cfk}&\text{Decay}&\text{RQM}& CLFQM\cite{klzdw}& PDG\cite{ParticleDataGroup:2024cfk}\\
			\hline
			$B^0\rightarrow \pi^- l^+\nu_l$&$13.4\pm 1.4$&&$15.0\pm0.5$&$B^+\rightarrow \pi^0 l^+\nu_l$ &$7.20\pm0.75$&$7.66\pm1.69$&$7.80\pm0.27$\\
	$B^0\rightarrow \pi^- \tau^+\nu_\tau$&$8.43\pm0.87$&&$<25$&$B^+\rightarrow \pi^0 \tau^+\nu_\tau$ &$4.54\pm0.47$&$5.21\pm1.15$\\	
		$B^0\rightarrow \rho^- l^+\nu_l$&$32.2\pm 3.4$&&$29.4\pm2.1$&$B^+\rightarrow \rho^0 e^+\nu_e$ &$17.4\pm1.8$&$21.3\pm4.7$&$15.8\pm1.1$\\
	$B^0\rightarrow \rho^- \tau^+\nu_\tau$&$18.0\pm1.9$&&&$B^+\rightarrow \rho^0 \tau^+\nu_\tau$ &$9.72\pm0.98$&$11.6\pm2.6$\\	
	$B^+\rightarrow \eta l^+\nu_l$&$3.40\pm 0.37$&$5.27\pm1.16$&$3.5\pm0.4$&$B^+\rightarrow \eta' l^+\nu_l$ &$2.62\pm0.28$&$2.56\pm0.56$&$2.4\pm0.7$\\
	$B^+\rightarrow \eta \tau^+\nu_\tau$&$2.20\pm0.25$&$3.23\pm0.71$&&$B^+\rightarrow \eta' \tau^+\nu_\tau$ &$1.51\pm0.17$&$1.36\pm0.30$\\	
	$B^+\rightarrow \omega l^+\nu_l$ &$14.1\pm1.5$&$20\pm4.4$&$11.9\pm0.9$
	&$B^+\rightarrow \omega \tau^+\nu_\tau$ &$7.91\pm0.80$&$10.7\pm2.4$\\	
	
		\end{tabular}\label{Brcompg}
	\end{ruledtabular}
\end{table}	

\subsection{Semileptonic $B$ decays to excited light mesons }

\subsubsection{Decays to radially excited mesons}

\begin{figure}
\centering
  \includegraphics[width=7.5cm]{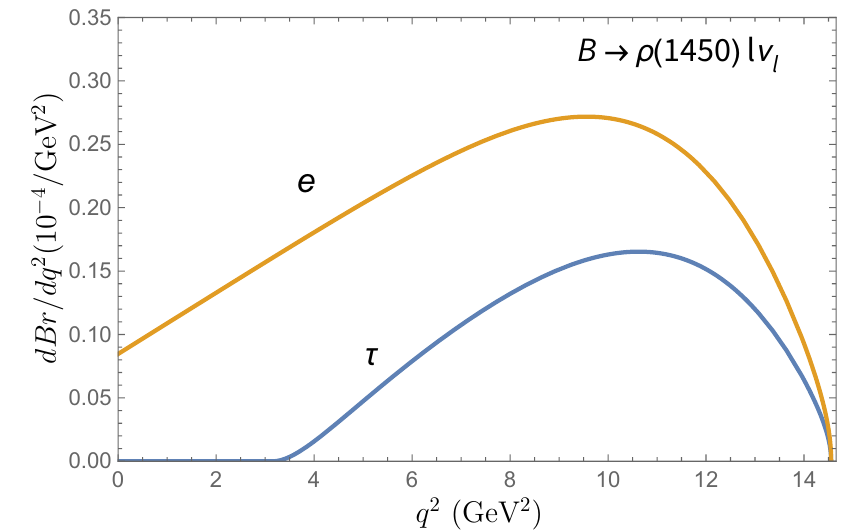} \quad \includegraphics[width=7.5cm]{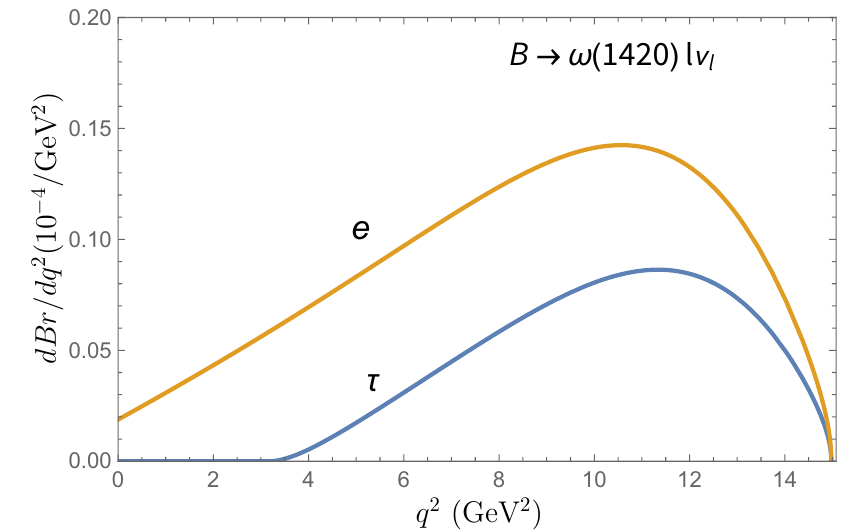}
\caption{Differential branching fractions of the semileptonic $B^0\rightarrow \rho^-(1450) l^+\nu_l$ and $B^+\rightarrow \omega(1420) l^+\nu_l$ decays.  }
\label{br2s}
\end{figure}

The obtained predictions for the semileptonic $B$ decays to radially excited light mesons are given in Table~\ref{Br1}. Branching fractions to both $2S$ and $3S$ states are given. It is seen that the branching fractions for decays to $3S$ states are almost an order of magnitude suppressed with respect to the branching fractions for decays to $2S$ states. The later ones have the same order as decays to the ground states. The largest branching fractions of order $10^{-4}$ are predicted for decays into the vector $\rho(1450)$ and $\omega(1420)$ mesons. Their differential branching fractions are plotted in Fig.~\ref{br2s}.

\begin{table}
	\caption{Branching fractions of the semileptonic $B$ decays to the radially excited ($2S$ and $3S$) states of light mesons ($\times 10^{-5}$).}
	\begin{ruledtabular}
		\begin{tabular}{cccc}			
			\text{Decay}&\text{Br}&\text{Decay}&\text{Br}\\
			\hline
			$B^0\rightarrow \pi^-(1300) e^+\nu_e$&$6.70^{+1.43}_{-1.22}$&$B^+\rightarrow \pi^0(1300) e^+\nu_e$ &$3.61^{+0.80}_{-0.65}$\\
			$B^0\rightarrow \pi^-(1300) \mu^+\nu_\mu$&$6.69^{+1.43}_{-1.22}$&$B^+\rightarrow \pi^0(1300) \mu^+\nu_\mu$ &$3.61^{+0.80}_{-0.65}$\\
	$B^0\rightarrow \pi^-(1300) \tau^+\nu_\tau$&$3.61^{+1.05}_{-0.82}$&$B^+\rightarrow \pi^0(1300) \tau^+\nu_\tau$ &$2.09^{+0.52}_{-0.43}$\\	
		$B^0\rightarrow \rho^-(1450) e^+\nu_e$&$28.5\pm 3.1$&$B^+\rightarrow \rho^0(1450) e^+\nu_e$ &$15.4\pm1.6$\\
			$B^0\rightarrow \rho^-(1450) \mu^+\nu_\mu$&$28.4\pm3.0$&$B^+\rightarrow \rho^0(1450) \mu^+\nu_\mu$ &$15.3\pm1.6$\\
	$B^0\rightarrow \rho^-(1450) \tau^+\nu_\tau$&$11.9\pm1.2$&$B^+\rightarrow \rho^0(1450) \tau^+\nu_\tau$ &$6.42\pm0.65$\\	
	$B^+\rightarrow \eta(1295) e^+\nu_e$&$3.66\pm 0.37$&$B^+\rightarrow \eta(1475) e^+\nu_e$ &$0.0012\pm0.0002$\\
			$B^+\rightarrow \eta(1295) \mu^+\nu_\mu$&$3.65\pm0.37$&$B^+\rightarrow \eta(1475) \mu^+\nu_\mu$ &$0.0012\pm0.0002$\\
	$B^+\rightarrow \eta(1295) \tau^+\nu_\tau$&$2.15\pm0.22$&$B^+\rightarrow \eta(1475) \tau^+\nu_\tau$ &$0.00057\pm0.00006$\\	
	$B^+\rightarrow \omega(1420) e^+\nu_e$ &$13.7\pm1.4$
			&$B^+\rightarrow \omega(1420) \mu^+\nu_\mu$ &$13.7\pm1.4$\\
	$B^+\rightarrow \omega(1420) \tau^+\nu_\tau$ &$6.10\pm0.65$\\	
	
	$B^0\rightarrow \pi^-(1800) e^+\nu_e$&$0.724\pm 0.075$&$B^+\rightarrow \pi^0(1300) e^+\nu_e$ &$0.391\pm0.040$\\
			$B^0\rightarrow \pi^-(1800) \mu^+\nu_\mu$&$0.721\pm0.074$&$B^+\rightarrow \pi^0(1800) \mu^+\nu_\mu$ &$0.389\pm0.040$\\
	$B^0\rightarrow \pi^-(1800) \tau^+\nu_\tau$&$0.165\pm0.017$&$B^+\rightarrow \pi^0(1800) \tau^+\nu_\tau$ &$0.089\pm0.09$\\	
		$B^0\rightarrow \rho^-(1900) e^+\nu_e$&$2.10\pm 0.22$&$B^+\rightarrow \rho^0(1450) e^+\nu_e$ &$1.13\pm0.12$\\
			$B^0\rightarrow \rho^-(1900) \mu^+\nu_\mu$&$2.09\pm0.22$&$B^+\rightarrow \rho^0(1900) \mu^+\nu_\mu$ &$1.13\pm1.2$\\
	$B^0\rightarrow \rho^-(1900) \tau^+\nu_\tau$&$0.590\pm0.060$&$B^+\rightarrow \rho^0(1900) \tau^+\nu_\tau$ &$0.318\pm0.0.032$\\	
	$B^+\rightarrow \eta(1760) e^+\nu_e$&$0.275\pm 0.028$&$B^+\rightarrow \omega(1960) e^+\nu_e$ &$1.16\pm0.12$\\
			$B^+\rightarrow \eta(1760) \mu^+\nu_\mu$&$0.274\pm0.028$&$B^+\rightarrow \omega(1960) \mu^+\nu_\mu$ &$1.15\pm0.12$\\
	$B^+\rightarrow \eta(1760) \tau^+\nu_\tau$&$0.087\pm0.009$&$B^+\rightarrow \omega(1960) \tau^+\nu_\tau$ &$0.281\pm0.029$\\	
	
		\end{tabular}\label{Br1}
	\end{ruledtabular}
\end{table}	

In Table~\ref{BrComr} we confront our results with flavor SU(3) predictions \cite{qsxw} which are available only for the semileptonic  $B$ decays into radially excited $2S$ light vector mesons. We see that theoretical results agree within { $(2-3)\sigma$} uncertainties.

\begin{table}
	\caption{Comparison of the calculated in RQM branching fractions with flavor SU(3) predictions \cite{qsxw} for the semileptonic  $B$ decays to the radially excited ($2S$) light vector mesons ($\times 10^{-5}$).}
	\begin{ruledtabular}
		\begin{tabular}{cc@{\!\!}c@{\!\!}ccc@{\!\!\!\!}c}
			Decay&RQM &flavor SU(3) \cite{qsxw}&Decay&RQM &flavor SU(3) \cite{qsxw}\\
			\hline
$B^0\rightarrow \rho^-(1450) e^+\nu_e$&$28.5\pm 3.1$&$19.26\pm2.87$&$B^+\rightarrow \rho^0(1450) e^+\nu_e$ &$15.4\pm1.6$&$10.36\pm1.52$\\	
$B^0\rightarrow \rho^-(1450) \tau^+\nu_\tau$&$11.9\pm1.2$&$8.23\pm1.34$&$B^+\rightarrow \rho^0(1450) \tau^+\nu_\tau$ &$6.42\pm0.65$&$4.43\pm0.71$\\
$B^+\rightarrow \omega(1420) e^+\nu_e$ &$13.7\pm1.4$&$11.34\pm2.20$&	$B^+\rightarrow \omega(1420) \tau^+\nu_\tau$ &$6.10\pm0.65$&$5.02\pm1.14$\\

\end{tabular}\label{BrComr}
\end{ruledtabular}
\end{table}

\subsubsection{Decays to orbitally excited mesons}

\begin{figure}
\centering
  \includegraphics[width=7.5cm]{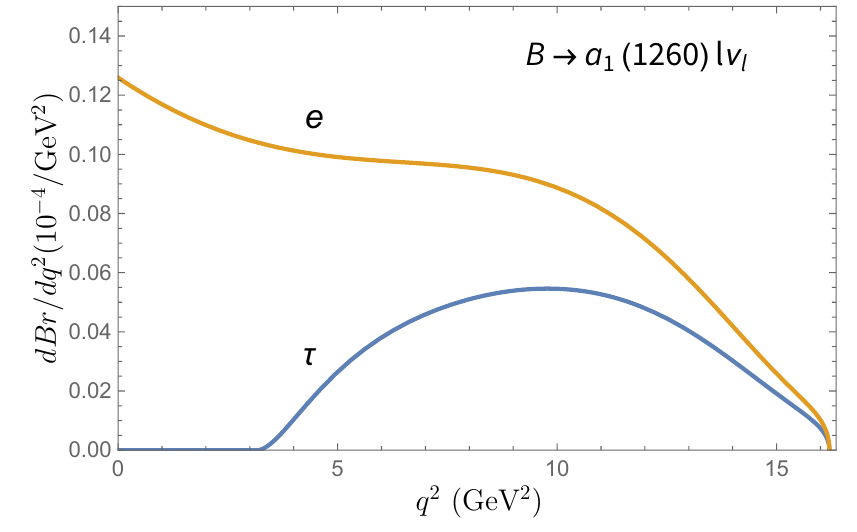} \quad \includegraphics[width=7.5cm]{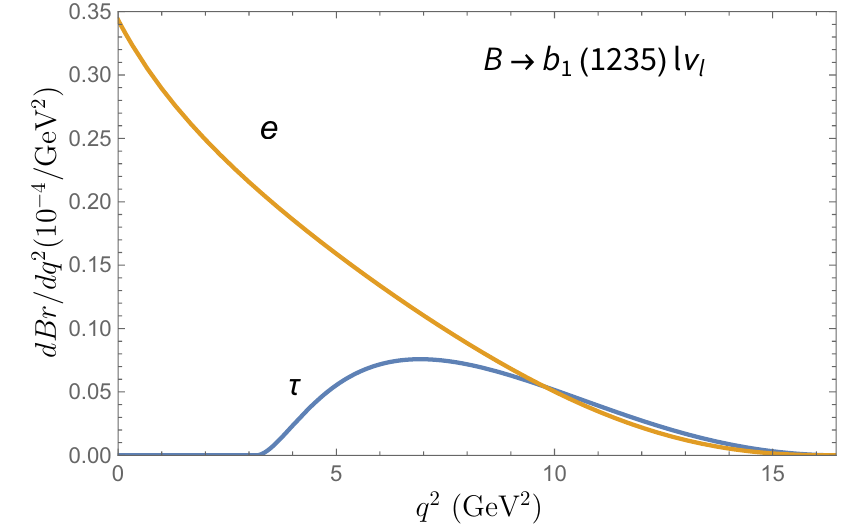}\\
  \includegraphics[width=7.5cm]{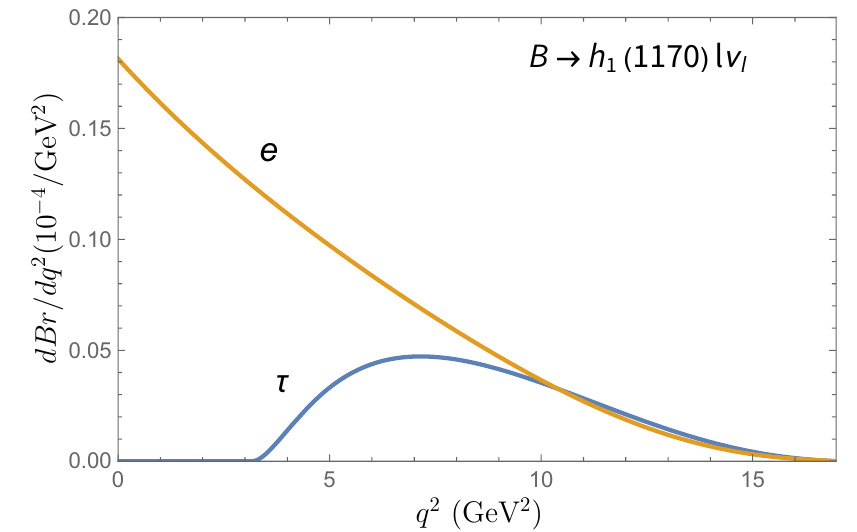} \quad \includegraphics[width=7.5cm]{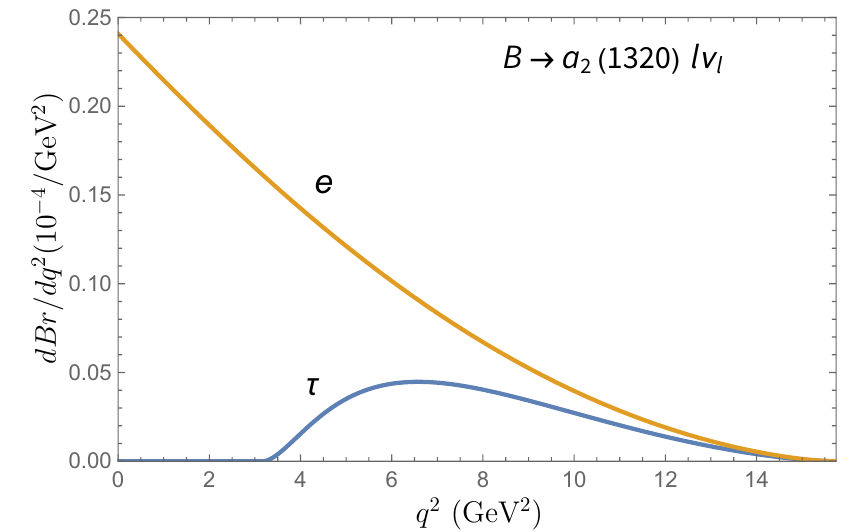}
\caption{Differential branching fractions of the semileptonic $B^0\rightarrow a_1^-(1260) l^+\nu_l$, $B^0\rightarrow b_1^-(1235) l^+\nu_l$, $B^+\rightarrow h_1(1170) l^+\nu_l$ and $B^0\rightarrow a_2^-(1320) l^+\nu_l$ decays.  }
\label{br1p}
\end{figure}

In Tables~\ref{Br1p1}-\ref{Br2p0} we present predictions of the RQM for the branching fractions of the semileptonic $B$ decays to the $1P$ and $2P$ states of light mesons. We separately list predictions for the decays to the isovector ($I=1$) and isoscalar ($I=0$) states. It is again found that for decays to the second orbital excitations ($2P$)  are suppressed by a factor of about 2 with respect to decays to the first orbital excitations ($1P$) with the only exception of the $B\to a_2(1700)l\nu_l$ decay, which is of the same magnitude as the $B\to a_2(1320)l\nu_l$ one.  The largest branching fractions of order $10^{-4}$ are predicted for $B^0$ decays to isovector $1P$ mesons and isovector $2P$ state $a_2(1700)$. In the isosclar sector only the decay $B^+\to h_1(1170)l^+\nu_l$ is predicted with the branching ratio of this order. We plot the branching fractions of these decays in Fig.~\ref{br1p}.

\begin{table}
	\caption{Branching fractions of the semileptonic $B$ decays to the orbitally excited ($1P$) states of light mesons with isospin $I=1$ ($\times 10^{-5}$).}
	\begin{ruledtabular}
		\begin{tabular}{cccc}			
			\text{Decay}&\text{Br}&\text{Decay}&\text{Br}\\
			\hline
			$B^0\rightarrow a_0^-(1450) e^+\nu_e$&$9.05\pm0.95$&$B^+\rightarrow a_0^0(1450) e^+\nu_e$ &$4.88\pm0.49$\\
			$B^0\rightarrow a_0^-(1450) \mu^+\nu_\mu$&$9.00\pm0.94$&$B^+\rightarrow a_0^0(1450) \mu^+\nu_\mu$ &$4.86\pm0.49$\\
	$B^0\rightarrow a_0^-(1450) \tau^+\nu_\tau$&$2.56\pm0.26$&$B^+\rightarrow a_0^0(1450) \tau^+\nu_\tau$ &$1.38\pm0.14$\\	
		$B^0\rightarrow a_1^-(1260) e^+\nu_e$&$14.44^{+1.75}_{-1.49}$&$B^+\rightarrow a_1^0(1260) e^+\nu_e$ &$7.80^{+1.05}_{-0.81}$\\
			$B^0\rightarrow a_1^-(1260) \mu^+\nu_\mu$&$14.38^{+1.73}_{-1.45}$&$B^+\rightarrow a_1^0(1260) \mu^+\nu_\mu$ &$7.76^{+1.00}_{-0.80}$\\
	$B^0\rightarrow a_1^-(1260) \tau^+\nu_\tau$&$5.17^{+0.80}_{-0.65}$&$B^+\rightarrow a_1^0(1260) \tau^+\nu_\tau$ &$2.79^{+0.40}_{-0.32}$\\	
	$B^0\rightarrow b_1^-(1235) e^+\nu_e$&$17.92\pm1.85$&$B^+\rightarrow b_1^0(1235) e^+\nu_e$ &$9.67\pm0.98$\\
			$B^0\rightarrow b_1^-(1235) \mu^+\nu_\mu$&$17.83\pm1.84$&$B^+\rightarrow b_1^0(1235) \mu^+\nu_\mu$ &$9.62\pm0.97$\\
	$B^0\rightarrow b_1^-(1235) \tau^+\nu_\tau$&$5.08\pm0.52$&$B^+\rightarrow b_1^0(1235) \tau^+\nu_\tau$ &$2.74\pm0.28$\\	
	$B^0\rightarrow a_2^-(1320) e^+\nu_e$&$13.61\pm1.41$&$B^+\rightarrow a_2^0(1320) e^+\nu_e$ &$7.35\pm0.74$\\
			$B^0\rightarrow a_2^-(1320) \mu^+\nu_\mu$&$13.52\pm1.40$&$B^+\rightarrow a_2^0(1320) \mu^+\nu_\mu$ &$7.30\pm0.73$\\
	$B^0\rightarrow a_2^-(1320) \tau^+\nu_\tau$&$2.79\pm0.29$&$B^+\rightarrow a_2^0(1320) \tau^+\nu_\tau$ &$1.51\pm0.16$\\	
		\end{tabular}\label{Br1p1}
	\end{ruledtabular}
\end{table}	

\begin{table}
	\caption{Branching fractions of the semileptonic $B$ decays to the orbitally excited ($1P$) states of light mesons with isospin $I=0$ ($\times 10^{-5}$). { The superscripts $^\ast$, $^\ddag$ and $^\dag$ correspond to the mixing schemes (\ref{eq:f0}), (\ref{f02}) and (\ref{eq:f0mix}), respectively.}}
	\begin{ruledtabular}
		\begin{tabular}{cccc}			
			\text{Decay}&\text{Br}&\text{Decay}&\text{Br}\\
			\hline
			$B^+\rightarrow f_0(1370) e^+\nu_e$&$\begin{array}{c}
			{3.06^{+0.65}_{-0.50}}^\ast \\{3.94^{+0.83}_{-0.64}}^\dag
			\end{array}$&$B^+\rightarrow f_0(1500) e^+\nu_e$ &$\begin{array}{c}
			{1.47\pm0.15}^\ast \\{0.93\pm0.09}^\dag
			\end{array}$\\
			$B^+\rightarrow f_0(1370) \mu^+\nu_\mu$&$\begin{array}{c}
			{3.04^{+0.65}_{-0.50}}^\ast \\{3.92^{+0.83}_{-0.64}}^\dag
			\end{array}$&$B^+\rightarrow f_0(1500) \mu^+\nu_\mu$ &$\begin{array}{c}
			{1.46\pm0.15}^\ast \\{0.92\pm0.09}^\dag
			\end{array}$\\
	$B^+\rightarrow f_0(1370) \tau^+\nu_\tau$&$\begin{array}{c}
			{0.96^{+0.37}_{-0.25}}^\ast \\{1.23^{+0.50}_{-0.33}}^\dag
			\end{array}$&$B^+\rightarrow f_0(1500) \tau^+\nu_\tau$ &$\begin{array}{c}
			{0.36\pm0.04}^\ast \\{0.23\pm0.03}^\dag
			\end{array}$\\
		$B^+\rightarrow f_0(1710) e^+\nu_e$&$0.404\pm0.042^\ddag$&$B^+\rightarrow f_0(1710) \mu^+\nu_\mu$ &$0.401\pm0.042^\ddag$\\
			$B^+\rightarrow f_0(1710) \tau^+\nu_\tau$&$0.075\pm0.008^\ddag$\\
		$B^+\rightarrow f_1(1285) e^+\nu_e$&$3.78\pm0.38$&$B^+\rightarrow f_1(1420) e^+\nu_e$ &$0.588\pm0.060$\\
			$B^+\rightarrow f_1(1285) \mu^+\nu_\mu$&$3.77\pm0.38$&$B^+\rightarrow f_1(1420) \mu^+\nu_\mu$ &$0.586\pm0.060$\\
	$B^+\rightarrow f_1(1285) \tau^+\nu_\tau$&$1.37\pm0.14$&$B^+\rightarrow f_1(1420) \tau^+\nu_\tau$ &$0.193\pm0.020$\\	
	$B^+\rightarrow h_1(1170) e^+\nu_e$&$10.95\pm1.98$&$B^+\rightarrow h_1(1415) e^+\nu_e$ &$0.0277\pm0.003$\\
			$B^+\rightarrow h_1(1170) \mu^+\nu_\mu$&$10.90\pm1.95$&$B^+\rightarrow h_1(1415) \mu^+\nu_\mu$ &$0.0276\pm0.003$\\
	$B+\rightarrow h_1(1170) \tau^+\nu_\tau$&$3.39\pm0.44$&$B^+\rightarrow h_1(1415) \tau^+\nu_\tau$ &$0.0065\pm0.0007$\\	
	$B^+\rightarrow f_2(1270) e^+\nu_e$&$6.36\pm0.64$&$B^+\rightarrow f'_2(1525) e^+\nu_e$ &$0.206\pm0.021$\\
			$B^+\rightarrow f_2(1270) \mu^+\nu_\mu$&$6.32\pm0.64$&$B^+\rightarrow f'_2(1525) \mu^+\nu_\mu$ &$0.204\pm0.021$\\
	$B^+\rightarrow f_2(1270) \tau^+\nu_\tau$&$1.31\pm0.13$&$B^+\rightarrow f'_2(1525) \tau^+\nu_\tau$ &$0.037\pm0.004$\\	
		\end{tabular}\label{Br1p0}
	\end{ruledtabular}
\end{table}

\begin{table}
	\caption{Branching fractions of the semileptonic $B$ decays to the orbitally and radially excited ($2P$) states of light mesons with isospin $I=1$ ($\times 10^{-5}$).}
	\begin{ruledtabular}
		\begin{tabular}{cccc}			
			\text{Decay}&\text{Br}&\text{Decay}&\text{Br}\\
			\hline
			$B^0\rightarrow a_0^-(1710) e^+\nu_e$&$2.20\pm0.25$&$B^+\rightarrow a_0^0(1710) e^+\nu_e$ &$1.18\pm0.15$\\
			$B^0\rightarrow a_0^-(1710) \mu^+\nu_\mu$&$2.18\pm0.24$&$B^+\rightarrow a_0^0(1710) \mu^+\nu_\mu$ &$1.18\pm0.15$\\
	$B^0\rightarrow a_0^-(1710) \tau^+\nu_\tau$&$0.530\pm0.056$&$B^+\rightarrow a_0^0(1710) \tau^+\nu_\tau$ &$0.286\pm0.030$\\	
		$B^0\rightarrow a_1^-(1640) e^+\nu_e$&$5.41\pm0.57$&$B^+\rightarrow a_1^0(1640) e^+\nu_e$ &$2.92\pm0.30$\\
			$B^0\rightarrow a_1^-(1640) \mu^+\nu_\mu$&$5.40\pm0.57$&$B^+\rightarrow a_1^0(1640) \mu^+\nu_\mu$ &$2.91\pm0.30$\\
	$B^0\rightarrow a_1^-(1640) \tau^+\nu_\tau$&$1.75\pm0.19$&$B^+\rightarrow a_1^0(1640) \tau^+\nu_\tau$ &$0.945\pm0.10$\\	
	$B^0\rightarrow b_1^-(1720) e^+\nu_e$&$3.21\pm0.34$&$B^+\rightarrow b_1^0(1720) e^+\nu_e$ &$1.73\pm0.20$\\
			$B^0\rightarrow b_1^-(1720) \mu^+\nu_\mu$&$3.20\pm0.34$&$B^+\rightarrow b_1^0(1720) \mu^+\nu_\mu$ &$1.73\pm0.20$\\
	$B^0\rightarrow b_1^-(1720) \tau^+\nu_\tau$&$0.899\pm0.092$&$B^+\rightarrow b_1^0(1720) \tau^+\nu_\tau$ &$0.485\pm0.051$\\	
	$B^0\rightarrow a_2^-(1700) e^+\nu_e$&$12.0\pm1.3$&$B^+\rightarrow a_2^0(1700) e^+\nu_e$ &$6.47\pm0.68$\\
			$B^0\rightarrow a_2^-(1700) \mu^+\nu_\mu$&$11.9\pm1.3$&$B^+\rightarrow a_2^0(1700) \mu^+\nu_\mu$ &$6.43\pm0.68$\\
	$B^0\rightarrow a_2^-(1700) \tau^+\nu_\tau$&$2.00\pm0.24$&$B^+\rightarrow a_2^0(1700) \tau^+\nu_\tau$ &$1.08\pm0.13$\\	
		\end{tabular}\label{Br2p1}
	\end{ruledtabular}
\end{table}	

\begin{table}
	\caption{Branching fractions of the semileptonic $B$ decays to the orbitally and radially excited ($2P$) states of light mesons with isospin $I=0$ ($\times 10^{-5}$). { The superscripts $^\ddag$ and $^\dag$ correspond to the mixing schemes (\ref{f02}) and (\ref{eq:f0mix}), respectively.}}
	\begin{ruledtabular}
		\begin{tabular}{cccc}			
			\text{Decay}&\text{Br}&\text{Decay}&\text{Br}\\
			\hline
$B^+\rightarrow f_0(1710) e^+\nu_e$&$1.54\pm0.16^\dag$&$B^+\rightarrow f_0(1770) e^+\nu_e$&$\begin{array}{c}
			{2.19\pm0.23}^\ddag \\{1.62\pm0.17}^\dag
			\end{array}$\\

$B^+\rightarrow f_0(1710) \mu^+\nu_\mu$ &$1.53\pm0.16^\dag$&$B^+\rightarrow f_0(1770) \mu^+\nu_\mu$&$\begin{array}{c}
			{2.17\pm0.23}^\ddag \\{1.61\pm0.17}^\dag
			\end{array}$\\
			$B^+\rightarrow f_0(1710) \tau^+\nu_\tau$&$0.324\pm0.004^\dag$&$B^+\rightarrow f_0(1770) \tau^+\nu_\tau$&$\begin{array}{c}
			{0.43\pm0.05}^\ddag \\{0.32\pm0.04}^\dag
			\end{array}$\\		
		
		$B^+\rightarrow f_1(1740) e^+\nu_e$&$2.42\pm0.26$&$B^+\rightarrow h_1(1650) e^+\nu_e$ &$2.84\pm0.31$\\
			$B^+\rightarrow f_1(1740) \mu^+\nu_\mu$&$2.41\pm0.26$&$B^+\rightarrow h_1(1650) \mu^+\nu_\mu$ &$2.82\pm0.31$\\
	$B^+\rightarrow f_1(1740) \tau^+\nu_\tau$&$0.83\pm0.09$&$B^+\rightarrow h_1(1650) \tau^+\nu_\tau$ &$0.63\pm0.07$\\	
	
	$B^+\rightarrow f_0(1855) e^+\nu_e$ &$1.47\pm0.16$&	$B^+\rightarrow f_2(1750) e^+\nu_e$&$2.39\pm0.25$\\
	$B^+\rightarrow f_0(1855) \mu^+\nu_\mu$ &$1.46\pm0.16$&$B^+\rightarrow f_2(1750) \mu^+\nu_\mu$ &$2.38\pm0.25$\\
	$B^+\rightarrow f_0(1855) \tau^+\nu_\tau$ &$0.26\pm0.03$&$B^+\rightarrow f_2(1750) \tau^+\nu_\tau$&$0.39\pm0.04$\\
		\end{tabular}\label{Br2p0}
	\end{ruledtabular}
\end{table}	

{ In Tables~\ref{Br1p0},~\ref{Br2p0} we present predictions for the branching fractions of the $B$ decays to isoscalar scalar light mesons ($1^3P_0$ and $2^3P_0$ states) in different mixing schemes (\ref{eq:f0})-(\ref{eq:f0mix}). The obtained results are marked by the corresponding superscripts ($^\ast,^\ddag,^\dag$). For decays involving the $f_0(1370)$, $f_0(1500)$ and $f_0(1770)$ we find that the branching ratios calculated in considered mixing schemes agree within $(2-3)\sigma$ uncertainties. Thus it will be very hard to distinguish these schemes in the semileptonic $B$ decays to these scalar mesons. On the other hand, for decays involving the $f_0(1710)$ we find a factor of about 3 difference between decay branching fractions. This is the consequence of the different interpretations of this meson. In the mixing scheme (\ref{eq:f0}) the $f_0(1710)$ corresponds to the $1^3P_0$ state with the largest admixture of the glueball and, thus, the corresponding decay is suppressed, while in the mixing scheme (\ref{eq:f0mix}) with the fragmented scalar glueball the $f_0(1710)$ is considered as  the $2^3P_0$ state with the small glueball admixture. Therefore the measurement of the $B^+\to f_0(1710)l^+\nu_l$ can distinguish between these mixing schemes.   }

In Table~\ref{Brcom1p} we compare our predictions for the branching ratios of the semileptonic $B$ decays to orbitally ($1P$) excited light mesons with other theoretical predictions. These decays were previously considered in the framework of the covariant light-front quark model (CLFQM) \cite{klzdw}, on the basis of the flavor SU(3) analysis \cite{qsxw}, in the perturbative QCD (PQCD) approach \cite{pqcd1,pqcd2,pqcd3} and light cone sum rules (LCSR) \cite{lcsr1,lcsr2,lcsr3}. { Since in these papers decays to the scalar $f_0$ mesons were calculated in the mixing scheme (\ref{eq:f0}), we give our predictions for the $B^+\to f_0l^+\nu_l$ branching fraction only in this scheme. } It is found that RQM results agree with almost all CLFQM, PQCD and LCSR predictions within error bars.
Contrary, the predictions based on the SU(3) flavor analysis are significantly different. They are about an order of magnitude larger than other results for the branching fractions of the semileptonic $B$ decays to axial vector mesons, while they are about a factor of 2 smaller for decays to tensor mesons. The origin of such discrepancy is not clear. Note that, as it was already mentioned above, we find reasonable agreement between the RQM and  SU(3) flavor results for radially excited light mesons (see Table~\ref{BrComr}).

\begin{table}
	\caption{Comparison of the calculated in RQM branching fractions with other theoretical predictions for the semileptonic $B$ meson decays to the orbitally ($1P$) excited light mesons ($\times 10^{-5}$).}
	\begin{ruledtabular}
		\begin{tabular}{@{}ccc@{}ccccccc@{}}
			\text{Decay}&\text{RQM}&CLFQM\cite{klzdw}&flavor SU(3)\cite{qsxw}&PQCD\cite{pqcd1,pqcd2,pqcd3}&LCSR\cite{lcsr1,lcsr2,lcsr3}\\
			\hline
	$B^0\rightarrow a_0^-(1450) e^+\nu_e$&$9.05\pm0.95$&&&$32.5^{+23.6}_{-13.6}$&$18^{+9}_{-6}$\\		
	$B^0\rightarrow a_0^-(1450) \tau^+\nu_\tau$&$2.56\pm0.26$&&&$13.2^{+9.7}_{-5.7}$&$6.3^{+3.4}_{-2.5}$\\	
	$B^+\rightarrow a_0^0(1450) e^+\nu_e$ &$4.88\pm0.49$&$2.72\pm0.6$\\
	$B^+\rightarrow a_0^0(1450) \tau^+\nu_\tau$ &$1.38\pm0.14$&$1.04\pm0.23$\\
	$B^+\rightarrow f_0(1370) e^+\nu_e$&$3.06^{+0.65}_{-0.50}$&&&$15.5^{+15.3}_{-6.5}$	\\
	$B^+\rightarrow f_0(1370) \tau^+\nu_\tau$&$0.96^{+0.37}_{-0.25}$&&	&$6.7^{+6.8}_{-2.9}$	\\	
$B^+\rightarrow f_0(1500) e^+\nu_e$ &$1.47\pm0.15$&$0.77\pm0.17$\\	
$B^+\rightarrow f_0(1500) \tau^+\nu_\tau$ &$0.36\pm0.04$&$0.29\pm0.06$\\	
$B^+\rightarrow f_0(1710) e^+\nu_e$&$0.404\pm0.042$&$0.21\pm0.05$\\
$B^+\rightarrow f_0(1710) \tau^+\nu_\tau$&$0.075\pm0.008$&$0.07\pm0.02$\\
$B^0\rightarrow a_1^-(1260) e^+\nu_e$&$14.44^{+1.75}_{-1.49}$&&$303.1\pm73.9$&$29.6^{+17.4}_{-13.9}$&$30.2^{+10.3}_{-10.3}$\\
$B^0\rightarrow a_1^-(1260) \tau^+\nu_\tau$&$5.17^{+0.80}_{-0.65}$&&$128.5\pm34.7$&$13.4^{+7.8}_{-6.3}$\\
$B^+\rightarrow a_1^0(1260) e^+\nu_e$ &$7.80^{+1.05}_{-0.81}$&$6.3\pm1.4$&$163.3\pm40.1$&&$32.4^{+13.3}_{-11.3}$\\
$B^+\rightarrow a_1^0(1260) \tau^+\nu_\tau$ &$2.79^{+0.40}_{-0.32}$&$2.68\pm0.59$&$69.1\pm18.7$\\
$B^+\rightarrow f_1(1285) e^+\nu_e$&$3.78\pm0.38$&$\{3.8,7.3\}$&$\{0.00832,156.7\}$&$6.9^{+3.7}_{-2.9}$&$17.5^{+6.5+0.4}_{-5.5-5.2}$\\
$B^+\rightarrow f_1(1285) \tau^+\nu_\tau$&$1.37\pm0.14$&$\{1.56,3.05\}$&$\{0.00335,64.3\}$\\
$B^+\rightarrow f_1(1420) e^+\nu_e$ &$0.588\pm0.060$&$\{0.17,1.34\}$&$96.0\pm79.9$&$0.4^{+0.2}_{-0.2}$&$0.6^{+0.3+4.4}_{-0.3-0.4}$\\
$B^+\rightarrow f_1(1420) \tau^+\nu_\tau$ &$0.193\pm0.020$&$\{0.06,0.52\}$&$38.7\pm32.3$\\
$B^0\rightarrow b_1^-(1235) e^+\nu_e$&$17.92\pm1.85$&&$522.1\pm185.7$&$28.8^{+15.1}_{-12.2}$&$19.3^{+8.4}_{-6.8}$\\
$B^0\rightarrow b_1^-(1235) \tau^+\nu_\tau$&$5.08\pm0.52$&&$246.3\pm82.3$\\
$B^+\rightarrow b_1^0(1235) e^+\nu_e$ &$9.67\pm0.98$&$7.7\pm1.7$&$281.6\pm100.1$\\
$B^+\rightarrow b_1^0(1235) \tau^+\nu_\tau$ &$2.74\pm0.28$&$3.0\pm0.7$&$132.6\pm44.2$\\
$B^+\rightarrow h_1(1170) e^+\nu_e$&$10.95\pm1.98$&$\{6.7,10.8\}$&$308.7\pm116.5$&$19.4^{+10.2}_{-8.2}$&$13.3^{+6.0+0.6}_{-4.9-2.7}$\\
$B+\rightarrow h_1(1170) \tau^+\nu_\tau$&$3.39\pm0.44$&$\{2.57,4.12\}$&$147.9\pm24.5$\\
$B^+\rightarrow h_1(1415) e^+\nu_e$ &$0.0277\pm0.003$&$\{0,0.16\}$&$\{5.37\times10^{-7},58.2\}$&$0.2^{+0.1}_{-0.1}$&$0.4^{+0.2+2.0}_{-0.1}$\\
$B^+\rightarrow h_1(1415) \tau^+\nu_\tau$ &$0.0065\pm0.0007$&$\{0,0.06\}$&$\{2.36\times10^{-7},24.5\}$\\	
$B^0\rightarrow a_2^-(1320) e^+\nu_e$&$13.61\pm1.41$&&$5.41\pm2.79$&$11.6^{+8.1}_{-5.7}$&16\\
$B^0\rightarrow a_2^-(1320) \tau^+\nu_\tau$&$2.79\pm0.29$&&$1.72\pm1.07$&$4.1^{+2.9}_{-2.0}$&6\\
$B^+\rightarrow a_2^0(1320) e^+\nu_e$ &$7.35\pm0.74$&&$2.92\pm1.51$\\
$B^+\rightarrow a_2^0(1320) \tau^+\nu_\tau$ &$1.51\pm0.16$&&$0.93\pm0.58$\\
$B^+\rightarrow f_2(1270) e^+\nu_e$&$6.36\pm0.64$&&$3.21\pm1.65$&$6.9^{+4.8}_{-3.4}$&8.5\\
$B^+\rightarrow f_2(1270) \tau^+\nu_\tau$&$1.31\pm0.13$&&$1.06\pm0.66$&$2.5^{+1.8}_{-1.3}$&3.4\\
$B^+\rightarrow f'_2(1525) e^+\nu_e$ &$0.206\pm0.021$&&$0.0456\pm0.0303$\\
$B^+\rightarrow f'_2(1525) \tau^+\nu_\tau$ &$0.037\pm0.004$&&$0.0127\pm0.0091$\\

		\end{tabular}\label{Brcom1p}
	\end{ruledtabular}
\end{table}

{ We can use our predictions for the branching fractions of the semileptonic $B$ decays to excited light mesons to identify channels which can be most probably measured in the near future. In Table~\ref{Rpi} we present the largest decay rates to the radially and orbitally excited meson channels normalized by the $B\to\pi l\nu_l$ rate. We define the ratio
\[R(M)=\frac{{\rm Br}(B\to M l\nu_l)}{{\rm Br}(B\to \pi l\nu_l)}. \] The ratios larger than one half are given.
In this ratio uncertainties from the CKM matrix element $|V_{ub}|$ cancel.

\begin{table}
	\caption{Ratios $R(M)$ for the most probable semileptonic $B$ decays to excited light mesons. }
	\begin{ruledtabular}
		\begin{tabular}{cccc}
			Decay&$R(M)$&Decay&$R(M)$\\	
			\hline
$B\to\rho(1450)$& $2.13\pm0.29$&$B\to\omega(1420)$&$1.90\pm0.26$\\
$B\to h_1(1170)$& $1.51\pm0.29$& $B\to b_1(1235)$&$1.34\pm0.18$\\
$B\to a_1(1260)$& $1.08^{+0.17}_{-0.15}$& $B\to a_2(1320)$&$1.01\pm0.14$\\
$B\to a_2(1700)$& $0.89\pm 0.12$& $B\to f_2(1270)$&$0.88\pm0.12$\\
$B\to a_0(1450)$&$0.68\pm0.09$&$B\to f_1(1285)$&$0.52\pm0.08$\\
		\end{tabular}\label{Rpi}
	\end{ruledtabular}
\end{table}
   
}

 In Table~\ref{Ratio} we compare our results for the ratios of the branching fractions of the $B$ semileptonic decay involving $\tau$ and $\mu$ with the ones predicted by CLFQM \cite{klzdw}
\begin{equation}
R_F=\dfrac{\Gamma(B\to F\tau^+\nu_\mu)}{\Gamma(B\to F\mu^+\nu_e)},
\end{equation}
where $F$ is the final light meson. Such comparison provides the test of the lepton universality.  If future experimental measurements find significant deviations from the presented values, this will signal the presence of the so-called new physics beyond the standard model. From this table we see that the RQM and CLFQM results are consistent with each other.

\begin{table}
	\caption{Ratio of the branching fractions $R$ with $\tau$ and $\mu$ in RQM and CLFQM.}
	\begin{ruledtabular}
		\begin{tabular}{cccccc}
			Decay&RQM&CLFQM\cite{klzdw}&Decay&RQM&CLFQM\cite{klzdw}\\	
			\hline
$B\to \pi$&$0.63\pm0.09$&$0.68\pm0.21$&$B\to a_1(1260)$&$0.36^{+0.06}_{-0.07}$&$0.43\pm0.13$\\
$B\to \rho$&$0.56\pm0.08$&$0.54\pm0.17$&$B\to f_1(1285)$&$0.36\pm0.05$&$\{0.21,0.80\}$\\
$B\to \eta$&$0.65\pm0.10$&$0.61\pm0.19$&$B\to f_1(1420)$&$0.33\pm0.05$&$\{0.05,3.0\}$\\
$B\to \eta'$&$0.57\pm0.08$&$0.53\pm0.17$&$B\to b_1(1235)$&$0.28\pm0.04$&$0.39\pm0.13$\\$B\to \omega$&$0.56\pm0.08$&$0.54\pm0.17$&$B\to h_1(1170)$&$0.31\pm0.07$&$\{0.24,0.61\}$\\
$B\to a_0(1450)$&$0.28\pm0.04$&$0.38\pm0.12$&$B\to h_1(1415)$&$0.23\pm0.04$&$\{0,0.38\}$\\
$B\to f_0(1370)$&$0.32^{+0.10}_{-0.14}$&&$B\to a_2(1320)$&$0.21\pm0.03$&\\
$B\to f_0(1500)$&$0.25\pm0.04$&$0.38\pm0.11$&$B\to f_2(1270)$&$0.21\pm0.03$&\\
$B\to f_0(1710)$&$0.19\pm0.03$&$0.33\pm0.12$&$B\to f'_2(1525)$&$0.18\pm0.03$&\\
		\end{tabular}\label{Ratio}
	\end{ruledtabular}
\end{table}

\subsection{Asymmetries and polarization characteristics of the semileptonic $B$ decays}

To complete our analysis of the semileptonic $B$ meson decays to the ground  excited states of light mesons  we calculate the forward-backward asymmetry $A_{FB}(q^2)$ defined in (\ref{Afb}), the lepton-side convexity parameter $C^l_F(q^2)$ (\ref{Clf}), the longitudinal $P^l_L(q^2)$ (\ref{Pll}) and transverse $P^l_T(q^2)$ (\ref{Plt}) polarizations of the final charged lepton, and longitudinal polarization $F_L(q^2)$ (\ref{Fl}) of the final vector meson.

In Tables~\ref{ObS}--\ref{Ob2P} we present our predictions for the mean values of the polarization and asymmetry parameters for the semileptonic $B$  decays. These values were obtained by separately integrating corresponding partial differential decay rates in numerators and the total decay rates in denominators. As it was argued previously \cite{Faustov2022} these observables are more sensitive to the quark dynamics in mesons than the total decay branching ratios. Thus, if experimentally measured, they can be used to discriminate between different theoretical approaches. Note that, while the decay branching ratios for decays involving $e^+$ and $\mu^+$ differ insignificantly, the polarization and asymmetry parameters for such decays have distinguishable values.

\begin{table}
\vspace*{-0.8cm}		\caption{Predictions for the asymmetry and polarization parameters for the semileptonic $B$ decays to the ground state and radially excited  light mesons for the positively charged leptons.}
	\begin{ruledtabular}
		\begin{tabular}{cccccc}
			\text{Decay}& $\langle A_{FB}\rangle $& $\langle C^l_F\rangle $ &$\langle P_L\rangle $&$\langle P_T\rangle $&$\langle F_L\rangle $\\
			\hline
$B\rightarrow \pi e^+\nu_e$&$-2.8\times 10^{-7}$&$-1.5$&$1$&$-0.0005$&\\
$B\rightarrow \pi \mu^+\nu_\mu$&$-0.004$&$-1.49$&$0.989$&$-0.009$&\\	
$B\rightarrow \pi \tau^+\nu_\tau$&$-0.220$&$-0.823$&$0.421$&$-0.723$&\\

$B\rightarrow \eta e^+\nu_e$&$-3.6\times 10^{-7}$&$-1.5$&$1$&$-0.0006$&\\
$B\rightarrow \eta \mu^+\nu_\mu$&$-0.006$&$-1.48$&$0.984$&$-0.120$&\\	
$B\rightarrow \eta \tau^+\nu_\tau$&$-0.280$&$-0.630$&$0.149$&$-0.853$&\\

$B\rightarrow \eta' e^+\nu_e$&$-5.3\times 10^{-7}$&$-1.5$&$1$&$-0.0008$&\\
$B\rightarrow \eta' \mu^+\nu_\mu$&$-0.008$&$-1.48$&$0.978$&$-0.150$&\\	
$B\rightarrow \eta' \tau^+\nu_\tau$&$-0.317$&$-0.467$&$-0.072$&$-0.882$&\\

$B\rightarrow \rho e^+\nu_e$&$-0.506$&$0.043$&$1$&$-0.0006$&$0.314$\\
$B\rightarrow \rho \mu^+\nu_\mu$&$-0.508$&$0.047$&$0.994$&$-0.011$&$0.314$\\	
$B\rightarrow \rho \tau^+\nu_\tau$&$-0.536$&$0.137$&$0.602$&$0.095$&$0.314$\\

$B\rightarrow \omega e^+\nu_e$&$-0.511$&$0.064$&$1$&$-0.0005$&$0.305$\\
$B\rightarrow \omega \mu^+\nu_\mu$&$-0.513$&$0.069$&$0.994$&$-0.008$&$0.305$\\	
$B\rightarrow \omega \tau^+\nu_\tau$&$-0.539$&$0.146$&$0.617$&$0.113$&$0.302$\\

$B\rightarrow \pi(1300) e^+\nu_e$&$-2.6\times 10^{-7}$&$-1.5$&$1$&$-0.0006$&\\
$B\rightarrow \pi(1300)\mu^+\nu_\mu$&$-0.005$&$-1.49$&$0.987$&$-0.111$&\\	
$B\rightarrow \pi(1300) \tau^+\nu_\tau$&$-0.300$&$-0.595$&$0.172$&$-0.823$&\\

$B\rightarrow \eta(1295) e^+\nu_e$&$-2.6\times 10^{-7}$&$-1.5$&$1$&$-0.0006$&\\
$B\rightarrow \eta(1295) \mu^+\nu_\mu$&$-0.005$&$-1.49$&$0.987$&$-0.111$&\\	
$B\rightarrow \eta(1295) \tau^+\nu_\tau$&$-0.302$&$-0.588$&$0.157$&$-0.831$&\\

$B\rightarrow \eta(1475) e^+\nu_e$&$-4.1\times 10^{-7}$&$-1.5$&$1$&$-0.0006$&\\
$B\rightarrow \eta(1475) \mu^+\nu_\mu$&$-0.007$&$-1.48$&$0.983$&$-0.125$&\\	
$B\rightarrow \eta(1475) \tau^+\nu_\tau$&$-0.312$&$-0.561$&$0.159$&$-0.814$&\\

$B\rightarrow \rho(1450) e^+\nu_e$&$-0.441$&$-0.131$&$1$&$-0.0001$&$0.392$\\
$B\rightarrow \rho(1450) \mu^+\nu_\mu$&$-0.444$&$-0.124$&$0.992$&$-0.022$&$0.392$\\	
$B\rightarrow \rho(1450) \tau^+\nu_\tau$&$-0.501$&$0.032$&$0.517$&$0.037$&$0.391$\\

$B\rightarrow \omega(1420) e^+\nu_e$&$-0.473$&$0.0041$&$1$&$-0.0001$&$0.332$\\
$B\rightarrow \omega(1420) \mu^+\nu_\mu$&$-0.475$&$0.007$&$0.995$&$-0.0002$&$0.332$\\	
$B\rightarrow \omega(1420) \tau^+\nu_\tau$&$-0.495$&$0.050$&$0.628$&$0.156$&$0.334$\\

$B\rightarrow \pi(1800) e^+\nu_e$&$-6.5\times 10^{-7}$&$-1.5$&$1$&$-0.0008$&\\
$B\rightarrow \pi(1800)\mu^+\nu_\mu$&$-0.009$&$-1.47$&$0.976$&$-0.143$&\\	
$B\rightarrow \pi(1800) \tau^+\nu_\tau$&$-0.339$&$-0.475$&$0.136$&$-0.785$&\\

$B\rightarrow \eta(1760) e^+\nu_e$&$-7.0\times 10^{-7}$&$-1.5$&$1$&$-0.0008$&\\
$B\rightarrow \eta(1760) \mu^+\nu_\mu$&$-0.010$&$-1.47$&$0.973$&$-0.152$&\\	
$B\rightarrow \eta(1760) \tau^+\nu_\tau$&$-0.314$&$-0.532$&$0.181$&$-0.752$&\\

$B\rightarrow \rho(1900) e^+\nu_e$&$-0.579$&$0.687$&$1$&$0.0002$&$0.028$\\
$B\rightarrow \rho(1900) \mu^+\nu_\mu$&$-0.579$&$0.686$&$0.995$&$0.035$&$0.028$\\	
$B\rightarrow \rho(1900) \tau^+\nu_\tau$&$-0.444$&$0.300$&$0.485$&$0.372$&$0.110$\\

$B\rightarrow \omega(1960) e^+\nu_e$&$-0.542$&$0.652$&$1$&$0.0002$&$0.043$\\
$B\rightarrow \omega(1960) \mu^+\nu_\mu$&$-0.541$&$0.650$&$0.997$&$0.0038$&$0.043$\\	
$B\rightarrow \omega(1960) \tau^+\nu_\tau$&$-0.430$&$0.319$&$0.595$&$0.411$&$0.037$\\
			\end{tabular}\label{ObS}
	\end{ruledtabular}
\end{table}

\begin{table}
\caption{Predictions for the asymmetry and polarization parameters for the semileptonic $B$ decays to the orbitally ($1P$) excited light mesons for the positively charged leptons.}
	\begin{ruledtabular}
		\begin{tabular}{c c c c c c}
			\text{Decay}& $\langle A_{FB}\rangle $& $\langle C^l_F\rangle $ &$\langle P_L\rangle $&$\langle P_T\rangle $&$\langle F_L\rangle $\\
			\hline
$B\rightarrow a_0(1450) e^+\nu_e$&$-6.2\times 10^{-7}$&$-1.5$&$1$&$-0.0006$&\\
$B\rightarrow a_0(1450) \mu^+\nu_\mu$&$-0.008$&$-1.47$&$0.979$&$-0.0006$&\\	
$B\rightarrow a_0(1450) \tau^+\nu_\tau$&$-0.259$&$-0.636$&$0.382$&$-0.629$&\\
$B\rightarrow f_0(1370) e^+\nu_e$&$-5.9\times 10^{-7}$&$-1.5$&$1$&$-0.0007$&\\
$B\rightarrow f_0(1370) \mu^+\nu_\mu$&$-0.008$&$-1.48$&$0.980$&$-0.118$&\\	
$B\rightarrow f_0(1370) \tau^+\nu_\tau$&$-0.267$&$-0.642$&$0.366$&$-0.665$&\\		
$B\rightarrow f_0(1500) e^+\nu_e$&$-6.8\times 10^{-7}$&$-1.5$&$1$&$-0.0007$&\\
$B\rightarrow f_0(1500) \mu^+\nu_\mu$&$-0.009$&$-1.47$&$0.977$&$-0.121$&\\	
$B\rightarrow f_0(1500) \tau^+\nu_\tau$&$-0.232$&$-0.630$&$0.410$&$-0.537$&\\		
$B\rightarrow f_0(1710) e^+\nu_e$&$-7.6\times 10^{-7}$&$-1.5$&$1$&$-0.0007$&\\
$B\rightarrow f_0(1710) \mu^+\nu_\mu$&$-0.010$&$-1.47$&$0.975$&$-0.123$&\\	
$B\rightarrow f_0(1710) \tau^+\nu_\tau$&$-0.180$&$-0.590$&$0.413$&$-0.374$&\\		
$B\rightarrow a_1(1260) e^+\nu_e$&$0.143$&$-0.332$&$1$&$-0.0005$&$0.481$\\
$B\rightarrow a_1(1260) \mu^+\nu_\mu$&$0.136$&$-0.311$&$0.981$&$-0.100$&$0.480$\\	
$B\rightarrow a_1(1260) \tau^+\nu_\tau$&$-0.039$&$0.087$&$0.447$&$-0.418$&$0.382$\\

$B\rightarrow f_1(1285) e^+\nu_e$&$0.107$&$-0.451$&$1$&$-0.0005$&$0.534$\\
$B\rightarrow f_1(1285) \mu^+\nu_\mu$&$0.101$&$-0.434$&$0.983$&$-0.097$&$0.534$\\	
$B\rightarrow f_1(1285) \tau^+\nu_\tau$&$-0.106$&$-0.002$&$0.402$&$-0.452$&$0.470$\\
$B\rightarrow f_1(1420) e^+\nu_e$&$0.068$&$-0.479$&$1$&$-0.0005$&$0.546$\\
$B\rightarrow f_1(1420) \mu^+\nu_\mu$&$0.062$&$-0.462$&$0.983$&$-0.095$&$0.546$\\	
$B\rightarrow f_1(1420) \tau^+\nu_\tau$&$-0.157$&$-0.023$&$0.408$&$-0.406$&$0.479$\\
$B\rightarrow b_1(1235) e^+\nu_e$&$0.0009$&$-1.497$&$1$&$-0.0010$&$0.999$\\
$B\rightarrow b_1(1235) \mu^+\nu_\mu$&$-0.013$&$-1.455$&$0.962$&$-0.189$&$0.999$\\	
$B\rightarrow b_1(1235) \tau^+\nu_\tau$&$-0.356$&$-0.372$&$-0.141$&$-0.871$&$0.999$\\	
$B\rightarrow h_1(1170) e^+\nu_e$&$0.0018$&$-1.490$&$1$&$-0.0010$&$0.996$\\
$B\rightarrow h_1(1170) \mu^+\nu_\mu$&$-0.011$&$-1.452$&$0.965$&$-0.180$&$0.996$\\	
$B\rightarrow h_1(1170) \tau^+\nu_\tau$&$-0.347$&$-0.398$&$-0.090$&$-0.868$&$0.994$\\	
$B\rightarrow h_1(1415) e^+\nu_e$&$-0.0051$&$-1.460$&$1$&$-0.0011$&$0.982$\\
$B\rightarrow h_1(1415) \mu^+\nu_\mu$&$-0.020$&$-1.416$&$0.960$&$-0.196$&$0.982$\\	
$B\rightarrow h_1(1415) \tau^+\nu_\tau$&$-0.358$&$-0.260$&$-0.212$&$-0.797$&$0.957$\\		
	$B\rightarrow a_2(1320) e^+\nu_e$&$0.090$&$-1.212$&$1$&$-0.0009$&$0.872$\\
$B\rightarrow a_2(1320) \mu^+\nu_\mu$&$0.078$&$-1.174$&$0.966$&$-0.162$&$0.871$\\	
$B\rightarrow a_2(1320) \tau^+\nu_\tau$&$-0.158$&$-0.296$&$0.225$&$-0.751$&$0.799$\\	
$B\rightarrow f_2(1270) e^+\nu_e$&$0.091$&$-1.207$&$1$&$-0.0009$&$0.870$\\
$B\rightarrow f_2(1270) \mu^+\nu_\mu$&$0.079$&$-1.169$&$0.967$&$-0.162$&$0.869$\\	
$B\rightarrow f_2(1270) \tau^+\nu_\tau$&$-0.155$&$-0.296$&$0.232$&$-0.747$&$0.796$\\		
$B\rightarrow f_2'(1525) e^+\nu_e$&$0.093$&$-1.185$&$1$&$-0.0009$&$0.860$\\
$B\rightarrow f_2'(1525) \mu^+\nu_\mu$&$0.081$&$-1.146$&$0.966$&$-0.165$&$0.860$\\	
$B\rightarrow f_2'(1525) \tau^+\nu_\tau$&$-0.154$&$-0.265$&$0.223$&$-0.738$&$0.778$\\

			\end{tabular}\label{Ob1P}
	\end{ruledtabular}
\end{table}

\begin{table}
	\caption{Predictions for the asymmetry and polarization parameters for the semileptonic $B$ decays to the orbitally and radially ($2P$) excited light mesons for the positively charged leptons.}
	\begin{ruledtabular}
		\begin{tabular}{c c c c c c}
			\text{Decay}& $\langle A_{FB}\rangle $& $\langle C^l_F\rangle $ &$\langle P_L\rangle $&$\langle P_T\rangle $&$\langle F_L\rangle $\\
			\hline
$B\rightarrow a_0(1710) e^+\nu_e$&$-7.0\times 10^{-7}$&$-1.5$&$1$&$-0.0008$&\\
$B\rightarrow a_0(1710) \mu^+\nu_\mu$&$-0.010$&$-1.47$&$0.975$&$-0.141$&\\	
$B\rightarrow a_0(1710) \tau^+\nu_\tau$&$-0.325$&$-0.506$&$0.196$&$-0.764$&\\		
$B\rightarrow f_0(1770) e^+\nu_e$&$-7.6\times 10^{-7}$&$-1.5$&$1$&$-0.0007$&\\
$B\rightarrow f_0(1770) \mu^+\nu_\mu$&$-0.010$&$-1.47$&$0.974$&$-0.139$&\\	
$B\rightarrow f_0(1770) \tau^+\nu_\tau$&$-0.294$&$-0.538$&$0.303$&$-0.670$&\\		
$B\rightarrow f_0(1855) e^+\nu_e$&$-8.1\times 10^{-7}$&$-1.5$&$1$&$-0.0008$&\\
$B\rightarrow f_0(1855) \mu^+\nu_\mu$&$-0.011$&$-1.47$&$0.972$&$-0.142$&\\	
$B\rightarrow f_0(1855) \tau^+\nu_\tau$&$-0.294$&$-0.528$&$0.304$&$-0.660$&\\		
$B\rightarrow a_1(1640) e^+\nu_e$&$0.384$&$0.415$&$1$&$-0.0002$&$0.149$\\
$B\rightarrow a_1(1640) \mu^+\nu_\mu$&$0.383$&$0.414$&$0.996$&$-0.047$&$0.149$\\	
$B\rightarrow a_1(1640) \tau^+\nu_\tau$&$0.261$&$0.278$&$0.573$&$-0.424$&$0.138$\\	
$B\rightarrow f_1(1740) e^+\nu_e$&$0.234$&$0.233$&$1$&$-0.0002$&$0.230$\\
$B\rightarrow f_1(1740) \mu^+\nu_\mu$&$0.233$&$0.231$&$0.996$&$-0.043$&$0.231$\\	
$B\rightarrow f_1(1740) \tau^+\nu_\tau$&$0.0092$&$0.102$&$0.465$&$-0.401$&$0.324$\\	
$B\rightarrow b_1(1720) e^+\nu_e$&$-0.033$&$-0.878$&$1$&$-0.0008$&$0.723$\\
$B\rightarrow b_1(1720) \mu^+\nu_\mu$&$-0.044$&$-0.846$&$0.971$&$-0.145$&$0.723$\\	
$B\rightarrow b_1(1720) \tau^+\nu_\tau$&$-0.225$&$-0.052$&$0.075$&$-0.328$&$0.572$\\	
$B\rightarrow h_1(1650) e^+\nu_e$&$-0.036$&$-0.953$&$1$&$-0.0009$&$0.757$\\
$B\rightarrow h_1(1650) \mu^+\nu_\mu$&$-0.051$&$-0.908$&$0.960$&$-0.167$&$0.756$\\	
$B\rightarrow h_1(1650) \tau^+\nu_\tau$&$-0.184$&$0.093$&$0.273$&$-0.188$&$0.463$\\	

$B\rightarrow a_2(1700) e^+\nu_e$&$0.135$&$-1.07$&$1$&$-0.0008$&$0.810$\\
$B\rightarrow a_2(1700) \mu^+\nu_\mu$&$0.124$&$-1.04$&$0.969$&$-0.155$&$0.809$\\	
$B\rightarrow a_2(1700) \tau^+\nu_\tau$&$-0.097$&$-0.211$&$0.251$&$-0.734$&$0.722$\\
$B\rightarrow f_2(1750) e^+\nu_e$&$0.140$&$-1.04$&$1$&$-0.0008$&$0.797$\\
$B\rightarrow f_2(1750) \mu^+\nu_\mu$&$0.129$&$-1.01$&$0.970$&$-0.151$&$0.796$\\	
$B\rightarrow f_2(1750) \tau^+\nu_\tau$&$-0.076$&$-0.202$&$0.292$&$-0.703$&$0.698$\\
			\end{tabular}\label{Ob2P}
	\end{ruledtabular}
\end{table}

\section{Conclusion}
\label{sec:concl}

Semileptonic $B$ meson decays to the ground and excited states of light mesons are investigated in the framework of the relativistic quark model based on the quasipotential approach. Such decays are very important for the determination of the CKM matrix element $|V_{ub}|$ from available experimental data and clarifying the nature of excited light mesons. To achieve this goal, first, the mixing schemes of the isosinglet mesons are discussed. On the basis of the previous light meson mass calculations the experimental candidates for the orbitally and radially excited light mesons are determined. They are collected in Table~\ref{Mass1}. In the following experimental masses of these states are used for calculations.

The form factors parameterizing the weak decay matrix elements of the semileptonic $B$ decays to the ground, radially and orbitally excited light mesons are calculated in the quasipotential approach. They are expressed as the overlap integrals of the initial $B$ meson and final light meson wave functions. These wave functions are known from the previous meson mass spectra calculations. It is important to point out that our approach explicitly accounts for all relativistic corrections including contributions of the intermediate negative energy states and relativistic transformations of the meson wave functions from rest to the moving reference frame. As a result the momentum transfer squared, $q^2$, dependence of the form factors is determined in the whole range of the transferred momentum without any extrapolations or model assumptions. The convenient parameterizations of the $q^2$ dependence of the form factors are given.

These form factors and helicity formalism are applied for the calculation of the differential and total decay branching fractions. The calculated branching fractions and experimental data on the semileptonic $B$ decays to the ground state light mesons are used for the extraction of the $|V_{ub}|$ value. Both data for the total and partial decay rates are used. It is found that decays to the pseudoscalar mesons favor a slightly larger $|V_{ub}|$ value than decays to the vector ones. However they well agree within uncertainties. The  value of the  $|V_{ub}|$ obtained in RQM (\ref{eq:vub}) is in a good agreement with the value extracted from the analysis of the inclusive decays.  This value is used for the calculations of the branching ratios of the semileptonic $B$ decays to radially and orbitally excited light mesons. Decays to the $2S$, $3S$, $1P$ and $2P$ states are considered. It is found that several decays to the excited light mesons have branching fractions of the order $10^{-4}$. Thus these decays can be measured at $B$ factories. Such measurements will be very important for the clarifying the nature of the excited light meson states. The obtained results are compared with previous predictions \cite{klzdw,qsxw,pqcd1,pqcd2,pqcd3,lcsr1,lcsr2,lcsr3}. Note that our calculations are the most complete ones since they include both radial and orbital excitations as well as decays involving the $e^+$, $\mu^+$ and $\tau^+$ leptons. In general reasonable agreement between theoretical predictions is found, except $B$ decays to the axial vector mesons for which the flavor SU(3) analysis \cite{qsxw} predicts values an order of magnitude larger than other approaches including RQM.

The asymmetry and polarization parameters for the semileptonic $B$ decays to the ground and excited light mesons are also calculated. These parameters are significantly more sensitive to the theoretical approaches than the total decay rates. Their measurement can help discriminating among theoretical approaches and can serve as a valuable tool of the determination of the nature of excited light mesons.

\begin{acknowledgments}
We are grateful for the useful discussions to D. Ebert and Hai-Yang Cheng.
The work of Xian-Wei Kang was supported in part by the National
Natural Science Foundation of China under Project No.~12275023.
\end{acknowledgments}

\bibliography{Semileptonic_Bexc}

\end{document}